\documentclass[sigconf]{acmart}

\AtBeginDocument{%
  \providecommand\BibTeX{{%
    \normalfont B\kern-0.5em{\scshape i\kern-0.25em b}\kern-0.8em\TeX}}}

\setcopyright{acmcopyright}
\copyrightyear{2023}
\acmYear{2023}

\acmConference[CHI '23]{ACM Conference on Human Factors in Computing Systems}{April 23--28,
 2023}{Hamburg, Germany}
\acmBooktitle{CHI '23: ACM Conference on Human Factors in Computing Systems,
April 23--38, 2023, Hamburg, Germany} 
\acmPrice{15.00}
\acmISBN{978-1-4503-XXXX-X/18/06}

\usepackage{scalerel}
\usepackage{multirow}
\usepackage{enumitem}
\usepackage{listings}
\usepackage{xparse}
\usepackage[utf8]{inputenc}
\usepackage{xcolor}
\usepackage{colortbl}
\usepackage{makecell}
\usepackage{hyperref}
\usepackage[nameinlink]{cleveref}
\usepackage{array}

\newcommand{\rr}[1]{{{#1}}}

\begin{document}

\title[{User-Engaged} Algorithm Auditing in Industry Practice]{Understanding Practices, Challenges, and Opportunities for {User-Engaged} Algorithm Auditing in Industry Practice}


\author{Wesley Hanwen Deng}
\email{hanwend@cs.cmu.edu}
\affiliation{%
  \institution{Carnegie Mellon University}
  \streetaddress{5000 Forbes Ave}
  \city{Pittsburgh}
  \state{PA}
  \postcode{15213}
  \country{USA}
}

\author{Bill Boyuan Guo}
\email{boyuang@andrew.cmu.edu}
\affiliation{%
  \institution{Carnegie Mellon University}
  \streetaddress{5000 Forbes Ave}
  \city{Pittsburgh}
  \state{PA}
  \postcode{15213}
  \country{USA}
}

\author{Alicia DeVrio}
\email{adevos@cs.cmu.edu}
\affiliation{%
  \institution{Carnegie Mellon University}
  \streetaddress{5000 Forbes Ave}
  \city{Pittsburgh}
  \state{PA}
  \postcode{15213}
  \country{USA}
}

\author{Hong Shen}
\email{hongs@cs.cmu.edu}
\affiliation{%
  \institution{Carnegie Mellon University}
  \streetaddress{5000 Forbes Ave}
  \city{Pittsburgh}
  \state{PA}
  \postcode{15213}
  \country{USA}
}

\author{Motahhare Eslami}
\authornote{Both authors contributed equally to this research.}
\email{meslami@cs.cmu.edu}
\affiliation{%
  \institution{Carnegie Mellon University}
  \streetaddress{5000 Forbes Ave}
  \city{Pittsburgh}
  \state{PA}
  \postcode{15213}
  \country{USA}
}

\author{Kenneth Holstein}
\authornotemark[1]
\email{kjholste@cs.cmu.edu}
\affiliation{%
  \institution{Carnegie Mellon University}
  \streetaddress{5000 Forbes Ave}
  \city{Pittsburgh}
  \state{PA}
  \postcode{15213}
  \country{USA}
}

\renewcommand{\shortauthors}{Wesley Deng et al.}

\begin{abstract}
Recent years have seen growing interest among both researchers and practitioners in \rr{\textit{user-engaged}} approaches to algorithm auditing, which directly engage users in detecting problematic behaviors in algorithmic systems. However, we know little about industry practitioners' current practices and challenges around \rr{user-engaged} auditing, nor what opportunities exist for them to better leverage such approaches in practice. To investigate, we conducted a series of interviews and iterative co-design activities with practitioners who employ \rr{user-engaged} auditing approaches in their work. Our findings reveal several challenges practitioners face in appropriately recruiting and incentivizing user auditors, scaffolding user audits, and deriving actionable insights from \rr{user-engaged} audit reports. Furthermore, practitioners shared organizational obstacles to \rr{user-engaged} auditing, surfacing a complex relationship between practitioners and user auditors. Based on these findings, we discuss opportunities for future HCI research to help realize the potential (and mitigate risks) of \rr{user-engaged} auditing in industry practice.
\end{abstract}



\maketitle

\section{Introduction}
In recent years, algorithm audits have risen to prominence as an approach to uncover biased, discriminatory, or otherwise harmful behaviors in algorithmic systems \cite{buolamwini2018gender, Buolamwini2019hearing, cramer2018assessing, eslami2017biased, eslami2019opacity, raji2020closing, sweeney2013discrimination, zou2018ai, sandvig2014auditing, metaxa2021auditing}. Today, algorithm audits are typically conducted by small groups of experts such as industry practitioners, researchers, and activists \cite{metaxa2021auditing, sandvig2014auditing}. Although expert-led approaches have been highly impactful, they often suffer from major blindspots, and fail to detect critical issues. For example, expert-led audits can fail when those conducting the audit lack the relevant cultural knowledge and lived experience to recognize and know where to look for certain kinds of harmful algorithmic behaviors \cite{DeVos2022TowardUA,holstein2019improving,shen2021everyday, young2019toward}. 

To overcome limitations of current algorithm auditing techniques, researchers in HCI and AI have begun to explore the potential of more \rr{\textit{user-engaged}} approaches to algorithm auditing, which directly engage users of AI products and services in surfacing harmful algorithmic behaviors. Recent years have seen many cases in which users organically came together to uncover and raise awareness about harmful behaviors in algorithmic systems they use day-to-day, which had eluded detection by industry teams or other expert auditors \cite{shen2021everyday}. Inspired by these observations, researchers have begun to explore the design of systems that can leverage the power of everyday users and crowds to surface harmful algorithmic behaviors that might otherwise go undetected (e.g., \cite{attenberg2015beat, cabrera2021discovering, kiela2021dynabench, nushi2018towards, ochigame2021search, lam2022enduser}). The designs of existing research systems span a spectrum of user engagement, from more practitioner-led approaches---such as crowdsourcing workflows in which users' testing and auditing activities are more heavily guided and constrained by requesters---to more user-led approaches in which users take greater initiative in directing their own activities.

In parallel to these research efforts, several major technology companies have begun to experiment with approaches that engage users in auditing their AI products and services for problematic behaviors. For example, in 2021 Twitter introduced its first ``algorithmic bias bounty'' challenge to engage users in identifying harmful biases in its image cropping algorithm \cite{chowdhury2021introducing}. In another effort, Google launched the ``AI Test Kitchen,'' a web-based application that invites users to experiment with some of Google’s latest AI-based conversational agents, and to report any problematic behaviors they encounter \cite{AItest}.\rr{ More recently, inspired by Twitter's ``algorithmic bias bounty,'' OpenAI initiated a ``Feedback Contest'' to encourage users to ``provide feedback on problematic model outputs'' during their interactions with ChatGPT chatbot \cite{ChatGPT_Feedback}.}

Despite growing interest from industry, there remains a gulf between the academic research literature on \rr{user-engaged} auditing and current industry practice. In particular, we still know little about industry AI practitioners’ current practices and challenges around \rr{user-engaged} auditing, and what opportunities exist for them to better leverage such approaches in practice. To investigate, in this paper we explore the following research questions:
\begin{itemize}[font=\bfseries,
  align=left,topsep=3pt]
\item[RQ1] What are AI practitioners’ current motivations and practices around user engagement in auditing their AI products and services for problematic algorithmic behaviors?
\item[RQ2] What opportunities and challenges do practitioners envision for \rr{user-engaged} approaches to better support their algorithm auditing efforts? 
\end{itemize}
We conducted a two-stage study with 12 industry practitioners from 9 technology companies, all of whom have experimented with engaging users in auditing their AI systems for problematic algorithmic behaviors. We first conducted semi-structured interviews to understand practitioners’ current practices and challenges around engaging users in AI testing and auditing. We then conducted co-design activities, working with practitioners to iteratively co-design three design artifacts as a way to further probe challenges perceived by practitioners and opportunities to better support \rr{user-engaged} approaches to algorithm auditing in industry practice.

Overall, our participants shared three major motivations for engaging users in AI testing and auditing: understanding users’ subjective experiences of problematic machine behaviors, overcoming their teams' blindspots when auditing their products and services, and gathering evidence from users to help them advocate for fairness work within their organizations. Participants shared prior experiences engaging users on different scales, from individual user study sessions, to focus group workshops, to large-scale user feedback and crowdsourcing activities. However, in doing so, practitioners encountered various challenges in engaging users effectively. For instance, practitioners discussed challenges they faced in recruiting and incentivizing the ``right'' group of auditors for a given task, with relevant identities and lived experiences. Participants also discussed the difficulties in scaffolding users towards productive auditing strategies, without biasing them to simply replicate industry teams' own blindspots. Finally, practitioners discussed the challenges of quantification when deriving actionable insights from \rr{user-engaged} auditing reports: relying upon the majority vote runs the risk of masking the very biases an audit is intended to uncover. In addition, participants shared broader organizational obstacles to \rr{user-engaged} auditing, highlighting key tensions that arise in practice when involving users in algorithm auditing efforts such as potential PR risks, profit motives that work against protecting marginalized groups, and privacy and legal concerns.

As private companies increasingly experiment with \rr{user-engaged} approaches to algorithm auditing, HCI research has a critical role to play in shaping more effective and responsible practices. \rr{To this end, this work contributes: 

\begin{itemize}

\item
An in-depth understanding of industry practitioners' motivations, current practices, and challenges in effectively engaging users in testing and auditing AI products and services. Our findings shed light on the types of problems practitioners aim to address through user engagement around algorithm auditing, as well as the the ways practitioners navigate organizational tensions around user involvement in AI development processes.

\item
A set of design implications for user-engaged algorithm auditing, beyond standard considerations for human computation or user feedback systems.

\item
Insights into the complex relationship between user auditors and industry practitioners working on responsible AI, suggesting opportunities for future HCI research to help realize the potential (and mitigate risks) of user-engaged auditing in industry practice.
    
\end{itemize}

}
\section{Related Work}
\subsection{Understanding and supporting responsible AI practices in industry contexts}

In recent years, significant effort has been directed towards the development of approaches, guidelines, and tools to help industry practitioners audit their AI products and services for unfair, biased, or otherwise harmful algorithmic behaviors (e.g., \cite{bird2020fairlearn, bellamy2018ai, googlePAIR, AIX360API, FairlearnAPI, MSRchecklist, mitchell2019model}). Early work in this area has largely been guided by advances in academic research on AI fairness \cite{barocas2016big, raji2019actionable, abebe2020roles, kroll2016accountable, eubanks2018automating, angwin2016machine}. Yet in a series of interview studies and surveys with industry AI practitioners, Holstein et al. \cite{holstein2019improving} found that there were major disconnects between the tools offered by the research community, versus the actual on-the-ground needs of industry AI practitioners. To address such gaps, a growing line of research in HCI has focused on better understanding industry AI practitioners needs and designing to support responsible AI practices in industry. For example, studies from Madaio et al. \cite{madaio2021assessing} and Rakova et al. \cite{rakova2021responsible} investigated the organizational challenges and barriers that practitioners face in practice when attempting to build more responsible AI systems.

Meanwhile, to better support responsible AI practices, companies have been developing responsible AI guidelines such as People + AI guidebook \cite{googlePAIR}, trustworthy AI principles \cite{varshney2019trustworthy}, AI fairness checklists \cite{MSRchecklist}, and responsible AI toolkits such as AI Explainability 360 \cite{AIX360API} and Fairlearn \cite{FairlearnAPI}. However, recent HCI research has surfaced gaps between fairness toolkits' capabilities and practitioners' needs \cite{lee2020landscape, deng2022exploring, lee2020landscape,richardson2021towards}. For example, Kaur et al. \cite{kaur2020interpreting} found that AI practitioners often over-trust and misuse AI explainability toolkits. Other work from Lee et al. and Deng et al. identified misalignment between the designs of existing fairness toolkits versus practitioners’ actual desires and usage of these toolkits \cite{lee2020landscape, deng2022exploring, richardson2021towards}. In interviews with AI practitioners, these authors found that, beyond the functionality provided by current toolkits, practitioners desired tools that could help them bring in perspectives from relevant domain experts and users, in order to aid them in auditing their AI systems \cite{deng2022exploring}. In the next sections, we discuss emerging work that aims to harness the power of users in algorithm auditing.

\subsection{The power of users in algorithm auditing}

\rr{Metaxa et al. \cite{metaxa2021auditing} define an algorithm audit as ``a method of repeatedly querying an algorithm and observing its output to draw conclusions about the algorithm's opaque inner workings and possible external impact.''} A growing body of work in HCI, AI, and related communities has developed tools and processes to audit algorithmic systems for biased, discriminatory, or otherwise harmful behaviors (e.g., \cite{buolamwini2018gender,metaxa2021auditing,sandvig2014auditing}). Past work in algorithm auditing has uncovered harmful algorithmic behaviors across a wide range of algorithmic systems, from search engines to hiring algorithms to computer vision applications \cite{noble2018algorithms, asplund2020auditing, sweeney2013discrimination, prates2020assessing, buolamwini2018gender, hannak2014measuring}. 

Today, algorithm audits are typically conducted by small groups of experts such as industry practitioners, researchers, activists, and government agencies \cite{metaxa2021auditing}. However, such expert-led audits often fail to surface serious issues that everyday users of algorithmic systems are quickly able to detect once a system is deployed in the real world \cite{holstein2019improving,shen2021everyday}. For instance, this approach can fail when those conducting the audit lack the relevant cultural knowledge and lived experience to recognize and know where to look for certain kinds of harmful algorithmic behaviors \cite{DeVos2022TowardUA,holstein2019improving,shen2021everyday,young2019toward}. In addition, expert-led audits may fail to detect certain harmful algorithmic behaviors because these behaviors only arise—or are only recognized as harmful—when a system is used in particular context or in particular ways, which auditors may fail to anticipate \cite{cramer2018assessing,eslami2017careful,friedman1996bias,holstein2019improving,selbst2019fairness,shen2021everyday}.

Recent years have seen many real-world cases in which \textit{users} have uncovered and raised awareness around harmful algorithmic behaviors in systems they use day-to-day (e.g., search engines \cite{buolamwini2018gender}, online rating/review systems \cite{sweeney2013discrimination, eslami2017biased}, and machine translation systems \cite{prates2020assessing}) although expert auditors had failed to detect these issues. Shen et al. \cite{shen2021everyday} developed the concept of ``everyday algorithm auditing'' to describe how everyday users detect, understand, and interrogate problematic machine behaviors via their daily interactions with algorithmic systems. In the cases these authors reviewed, regular users of a wide range of algorithmic systems and platforms came together organically to hypothesize and test for potential biases. More recently, DeVos et al. \cite{DeVos2022TowardUA} conducted a series of behavioral studies to better understand how users are often able to be so effective, both individually and collectively, in surfacing harmful algorithmic behaviors that more formal or expert-led auditing approaches fail to detect. As discussed next, recent research is beginning to explore ways to harness the power users in algorithm auditing to overcome limitations of expert-led approaches.

\subsection{Supporting \rr{user-engaged} algorithm auditing}

Recognizing the power of users in algorithm auditing, researchers have begun to explore the design of systems to support more \textit{\rr{user-engaged}} approaches \cite{DeVos2022TowardUA, lam2022enduser} to algorithm auditing, which directly engage users in surfacing harmful algorithmic behaviors that might otherwise go undetected.

A line of work has developed interfaces, interactive visualizations, and crowdsourcing pipelines to support people in actively searching for algorithmic biases and harmful behaviors \cite{attenberg2015beat, cabrera2021discovering, kiela2021dynabench, nushi2018towards}. The designs of these research systems span a spectrum of user-engagement, from more practitioner-led approaches to more user-led approaches in which users take greater initiative and control in directing their efforts. For example, Ochigame and Ye developed a web-based tool called Search Atlas, which enables users to easily conduct side-by-side comparisons of the Google search results they might see if they were located in different countries to spot  \cite{ochigame2021search}. Kiela et al. developed a general research platform called Dynabench, which invites users to try to identify erroneous and potentially harmful behaviors in AI models \cite{kiela2021dynabench}. Using Dynabench, users can generate test inputs to a model to try to find problematic behaviors, flag behaviors they identify, and provide brief open-text responses if they wish to offer additional context. More recently, Lam et al. developed a tool called ``IndieLabel,” in order to empower end users to detect and flag potential algorithmic biases and then author audit reports to communicate these to relevant decision-makers \cite{lam2022enduser}.

In parallel, several major technology companies have begun to experiment with approaches that engage users in auditing their AI products and services for problematic behaviors. For example, in 2021 Twitter introduced its first ``algorithmic bias bounty'' challenge to engage users in identifying harmful biases in its image cropping algorithm \cite{chowdhury2021introducing}. In another effort, Meta adopted the Dynabench platform described above, to discover potentially harmful behaviors in natural language processing models \cite{kiela2021dynabench}. More recently, Google launched the ``AI Test Kitchen,'' a web-based application that invites users to experiment with Google’s latest LLMs-powered conversational agents, and to report any problematic behaviors they encounter, with the stated goal of engaging users in \emph{``learning, improving, and innovating responsibly on AI together''} \cite{AItest}. \rr{In addition, organizations like OpenAI and HuggingFace are beginning to include built-in interface features that invite users to report harmful algorithmic behaviors they encounter while interacting with LLM-powered applications like text-to-image generation tools. HuggingFace developed features to engage end users in flagging ethical/legal issues on their API \cite{HuggingFace}. In addition, OpenAI initiated a feedback contest around their LLM-based tool ChatGPT, with the goal of encouraging users to ``provide feedback on problematic model outputs'' \cite{ChatGPT_Feedback}.}

Despite growing interest in both academia and industry, there remains a gulf between the academic research literature on \rr{user-engaged} auditing and current industry practice. To date, little is known about industry AI practitioners’ current practices and challenges around \rr{user-engaged} auditing, nor what opportunities exist for them to better leverage such approaches in practice. In this paper, we take a first step towards understanding current practices, challenges, and design opportunities for \rr{user-engaged} approaches to algorithm auditing in industry practice.

\section{Method}

 \begin{figure*}[t]
  \centering
  \includegraphics[width=0.97\linewidth]{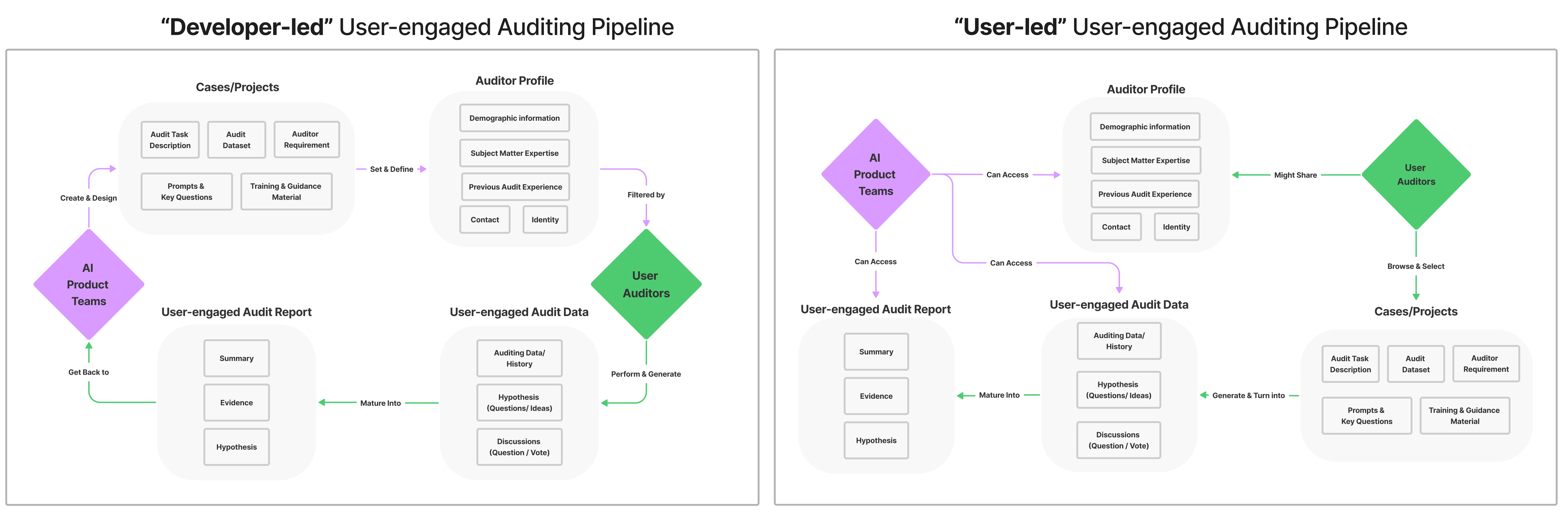}
  \caption{
    Two potential \rr{user-engaged} auditing pipeline designs that were iteratively co-designed with participants. The left image shows the ``developer-led’’ pipeline design, and the right image shows the ``user-led’’ pipeline design. Each figure illustrates a possible interaction flow between user auditors and AI product teams, showing how auditing tasks are created, how background information on user auditors is shared, how user auditing reports are generated based on auditors’ findings, and how these reports are shared with AI product teams. During the co-design activity, participants could zoom in, annotate, and modify the details. We used these pipeline flowcharts as \textbf{probes}, \textbf{not as final products}, to investigate more deeply on practitioners’ challenges and desires.
}
  \Description{Two images showing two potential \rr{user-engaged} auditing pipelines designs. The left image shows the ``developer-led’’ pipeline design, and the right image shows a ``user-led’’ pipeline design.}
  \label{fig:pipeline}
\end{figure*}

\subsection{Study design} \label{Methods}

We conducted a two-stage study involving semi-structured interviews followed by iterative co-design activities. We first conducted semi-structured interviews to understand participants’ current practices and challenges around engaging users in AI testing and auditing. In the next stage, we engaged participants in a co-design activity to further probe the opportunities and challenges in supporting \rr{user-engaged} algorithm auditing in industry practice. We worked with participants to iteratively design three artifacts: a \textbf{\rr{user-engaged} audit report}, representing a ``wish list'' of types of information that they would ideally want to solicit through a \rr{user-engaged} auditing approach, and \textbf{two \rr{user-engaged} auditing pipelines}, building upon initial designs informed by interview findings and insights from prior literature \cite{sandvig2014auditing, metaxa2021auditing, shen2021everyday, DeVos2022TowardUA}. Throughout the study, we iterated on these design artifacts based on feedback and design ideas from prior participants. We used these artifacts and the \emph{process} of co-designing them to provoke deeper conversations around participants' desires, as well as potential risks they anticipate, for new systems that support \rr{user-engaged} auditing. 

Following an iterative co-design process similar to prior work (cf. \cite{holstein2019co, madaio2020co}), for our first five participants, we ran the two stages of our study in separate sessions in order to better inform the design of the initial versions of the artifacts based on the needs and desires these participants expressed in the first set of interviews. However, we soon encountered difficulty in retaining industry participants due to their busy schedules (e.g., one participant was not able to return to complete the co-design). Therefore, after our first five participants, we ran both stages in a single session. We then continued to iterate on the artifacts during the study sessions themselves. Below, we describe each of these activities in more detail.\footnote{We also provide our interview and co-design protocol in the supplementary material. }

\subsubsection{Stage one: Semi-structured interviews}

To understand practitioners' current practices around engaging users in AI testing and auditing, we conducted semi-structured interviews, each lasting up to an hour. We adopted a directed storytelling approach \cite{evenson2006directed}. We first asked participants to describe their team's prior experiences in trying to detect or address biased or harmful behaviors in their AI products or services, with a specific focus on whether, why, and how they engaged users in the process. For example, we asked \textit{``Could you describe how your team attempted to engage users in auditing the AI products and services you mentioned''} and \textit{``What motivated you or your team to engage users in this way?''} Through follow-up questions, we probed deeper into challenges participants had encountered when attempting to engage users in the auditing process. As participants shared specific challenges they had encountered, we also invited them to share ideas for potential solutions to these challenges. For example, in response to specific challenges raised by participants, we asked \textit{``How did your team attempt to tackle these challenges?''} and \textit{``How effective were your team’s approaches?''}

\subsubsection{Stage two: Iterative co-design activities}

To further envision future opportunities and solicit potential challenges and risks for \rr{user-engaged} algorithm auditing approaches, following the interviews, we then involved participants in a series of co-design activities, following an iterative co-design process similar to prior work (cf. \cite{holstein2019co, madaio2020co}). This stage of the study lasted up to 45 minutes, and involved participants in co-design around three design artifacts: a \textbf{\rr{user-engaged} audit report} and two \textbf{\rr{user-engaged} audit pipeline flowcharts}. We first designed initial versions of these artifacts based on participant needs and desires expressed during stage one, as well as prior research on \rr{user-engaged} algorithm auditing \cite{shen2021everyday, DeVos2022TowardUA}. We then iterated on their designs with practitioners throughout the co-design activities. We note that these design artifacts were not the goal of our study, but rather served as tools to probe more deeply on practitioners’ challenges and desires. Below, we describe the process of co-designing these three artifacts, and how we used this process to probe on future opportunities and risks of \rr{user-engaged} audits. 

\textbf{\rr{User-engaged} audit report}: We invited each participant to contribute to the design of a report that they would ideally like to see as the \textit{output} of a \rr{user-engaged} auditing process. We first asked participants open-ended questions such as \textit{``What information would you ideally want the service to report back to your team?''} and encouraged them to sketch as they generated new ideas. To help participants come up with ideas, we presented participants with example of actual written responses generated by users during a \rr{user-engaged} algorithm auditing workshop in prior work\footnote{We provide the example user report we used in the study in the supplementary material} \cite{DeVos2022TowardUA}. Viewing examples of actual user responses on an auditing task provided an opportunity for participants to reflect upon gaps in their own report designs, and to anticipate potential challenges in soliciting useful information from user auditors (cf. \cite{holstein2020replay, madaio2020co}). We then presented participants with a \rr{user-engaged} audit report template (see Fig. \ref{fig:report} in Appendix) that we initially designed based on previous work and iterated through the design during the study. Our goal was to further probe participants' feedback by providing them with potential content that an audit report can consist of (such as information about the auditors, details of the reported issue, evidence, the severity of the issue, etc.). However, we intentionally showed this report after participants generated ideas about what a report can include to avoid biasing them towards a specific report format, yet giving them the opportunity to discuss other options and iterating through the report design based on their initial ideas.

\textbf{\rr{User-engaged} audit pipeline flowcharts}: We also co-designed two opposing caricatures of \rr{user-engaged} audit pipeline designs with participants \rr{which varied in the degree of initiative users assumed in the auditing process:}  (1) a \textbf{``developer-led''} pipeline, in which the audits were primarily initiated and coordinated by the developers of an AI product or service; (2) a \textbf{``user-led''} pipeline, in which the audits were primarily initiated and coordinated by users (see Fig. \ref{fig:pipeline}). For the developer-led pipeline, the AI product teams can fully control what AI systems (or what parts of the AI systems) should be audited, who should be considered eligible auditors, and how the auditing should be executed. In the ``user-led'' pipeline, users are those who initiate and control the auditing based on their interactions with the AI systems. Users could collectively initiate the auditor selection criteria, defining auditing protocol, generating audit data, and synthesizing reports. In this case, AI practitioners can only access the audit data without getting involved in or having a say in the auditing process. While we anticipated that \emph{neither} of these caricatures would represent ideal designs from industry practitioners’ perspectives, we presented these in order to \emph{provoke further discussion} about design trade-offs between greater user versus developer control in auditing processes.

\begin{table*}[]
\centering
\resizebox{\textwidth}{!}{%
\begin{tabular}
    {
    | >{\centering\arraybackslash}p{0.05\linewidth}
    | >{\centering\arraybackslash}p{0.1\linewidth}
    | >{\centering\arraybackslash}p{0.1\linewidth}
    | >{\centering\arraybackslash}p{0.2\linewidth}
    | >{\centering\arraybackslash}p{0.3\linewidth}
    | >{\centering\arraybackslash}p{0.1\linewidth}
    | >{\centering\arraybackslash}p{0.1\linewidth}
    |
    }
\hline
 &
  \textbf{Company size} &
  \textbf{Job title} &
  \textbf{Types of AI products and services} &
  \textbf{Experience with \rr{user-engaged} auditing} &
  \textbf{Years of AI fairness experience} &
  \textbf{Is fairness work part of their official role?} \\ \hline
P1 &
  1000–4,999 &
  Director and Product Lead &
  AI-powered knowledge graph for academic literature search &
  Engaging search users in assessing potential biases in the underlying knowledge graph. &
  4 &
  Yes \\ \hline
P2 &
  10–50 &
  Senior Director &
  ML model to predict potential donors &
  Engaging marginalized community members in surfacing potential biases in their team’s ML model &
  3 &
  No (self motivated) \\ \hline
P3 &
  25,000+ &
  ML Engineer &
  Natural language processing (NLP) applications &
  Engaging users in rating the risk of representational harms &
  2 &
  Yes \\ \hline
P4 &
  25,000+ &
  Senior Technical Lead &
  A diverse range of AI products built by their customers &
  Engaging users in auditing a range of AI products built by their customers &
  3 &
  Yes \\ \hline
P5 &
  25,000+ &
  Senior Product Manager &
  Sentiment analysis; OCR recognition &
  Leading several groups of AI product teams on engaging users in testing and auditing their AI products &
  5 &
  Yes \\ \hline
P6 &
  25,000+ &
  UX Researcher &
  NLP-powered products (e.g., conversational agents) &
   Engaging users in testing conversational agents for potentially harmful behaviors &
  2 &
  No (self motivated) \\ \hline
P7 &
  25,000+ &
  UX Researcher &
  Computer vision applications (e.g., image search) &
  Building an interface to engage users in auditing their image search engine &
  2 &
  Yes \\ \hline
P8 &
  25,000+ &
  ML Researcher &
  Information retrieval and image processing &
  Building internal crowdsourcing tools for AI auditing &
  3 &
  Yes \\ \hline
P9 &
  5,000 - 24,999 &
  Data Scientist &
  Recommendation system &
  Engaging users and impacted stakeholders in auditing their recommendation system  &
  3 &
  Yes \\ \hline
P10 &
  25,000+ &
  UX Researcher &
  Large language model-based conversational agent and generative image model &
  Building a web-based application to engage users in flagging the potential biased and harmful behavior in a conversational agent &
  2 &
  Yes \\ \hline
P11 &
  5,000 - 24,999 &
  Senior Researcher &
  Recommendation system &
  Leading research efforts and producing concrete organizational policy around \rr{user-engaged} algorithm auditing &
  2 &
  Yes \\ \hline
P12 &
  25,000+ &
  UX Researcher &
  NLP &
  Engaging marginalized communities in auditing biases in their NLP products for low-resource languages &
  3 &
  No (self motivated) \\ \hline
\end{tabular}%
}
\Description[A table of 12 research participants’ backgrounds, showing their industry roles and efforts around engaging users in auditing their AI products and services ]{Column one lists the participant number. Column two lists the size of the company each participant works at. The majority work at large corporations with more than 25000 employees. Column three lists participants’ titles at the company. Half work as either UX or ML researchers. Column four captures a diverse set of AI products and services each participant works on, including NLP applications, recommendation systems, and Computer Vision applications. Column five describes participants’ experience with \rr{user-engaged} auditing. Column six lists participants’ years of experience with AI fairness. All have a minimum of 2 years of experience with AI fairness. Column seven lists whether fairness-related work is part of each participant’s official responsibility. Nine out of twelve were officially responsible, and the rest were self-motivated.}
\caption{Summary of participants' backgrounds and relevant experience.}
\label{tab:participants}
\end{table*}

\subsection{Participants}
We adopted a purposive sampling approach \cite{campbell2020purposive}, with the aim of recruiting industry practitioners who either (1) had direct prior experience employing \textit{\rr{user-engaged} algorithm auditing} approaches, or (2) had an interest in such approaches and had adjacent experience \textit{crowdsourcing} approaches as part of their AI work. Specifically, using an online screening survey, we recruited members of industry teams that design and build AI products and services, and who had \textit{already} attempted to engage users in detecting fairness-related issues in their AI systems. In addition, we opened up the study to interested practitioners who had \textit{not} yet experimented with \rr{user-engaged} approaches to algorithm auditing, but who had prior experience using crowdsourcing approaches in other areas of their AI work. We broadened our criteria to include these participants because we expected that prior experience with crowdsourcing would help participants envision ways \rr{user-engaged} approaches might support their algorithm auditing efforts. In the end, however, \emph{all} 12 of our participants had direct prior experience experimenting with \rr{user-engaged} approaches to algorithm auditing (see Table \ref{tab:approaches}). In addition, all but three of our participants had prior experience with crowdsourcing methods. 

We recruited our participants through social media (e.g., Linkedin and Twitter), and through direct contacts at large technology companies. As discussed in prior literature that studies responsible AI practices in industry (e.g., \cite{holstein2019improving, lee2020landscape, deng2022exploring}), recruitment for such studies can be highly challenging. Practitioners are often wary of participating in such interview studies, for instance, given that participation may require admitting the existence of flaws in their products and services that have not been made public, or sharing disagreements about their companies' current organizational culture. Although we assured potential participants that their responses would be carefully de-identified, as discussed below, we expect that such concerns likely had an influence on our recruitment. 

In total, 25 practitioners completed the recruitment screening form, of which 18 met our recruitment criteria. Ultimately, 12 of these practitioners, spanning 9 companies, responded to our study invitation and participated in the study. All 12 participants participated in the interview session; all except one participant participated in the co-design session. (P4 was not able to return to complete the co-design activity due to busy schedule.)  All participants were compensated at a rate of \$35 per hour for their participation. Table \ref{tab:participants} overviews participants‘ job titles, their years of experience with \rr{user-engaged} auditing, their company size, the types of AI products or services they worked on, and their experiences with \rr{user-engaged} auditing. While 9 of our participants conducted \rr{user-engaged} algorithm auditing as part of their main job function; three participants (P2, P6, P12) engaged users in algorithm auditing on their own initiative, to help them in advocating for fairness issues to be addressed within their organizations.

Following prior work on responsible AI practices in industry \cite{holstein2019improving, madaio2020co, madaio2021assessing, deng2022exploring}, to avoid identifying individual participants who work at the forefront of sensitive topics, details about participants' demographics are omitted, and we abstract some details about participants' companies and roles. In addition, we assured participants that we would not ask them to reveal any confidential or personally identifying information about their colleagues and that we would de-identify all responses at the individual, team, and organization levels. Finally, participants were instructed that they were free to skip any questions they were uncomfortable answering, or to leave the session at any time for any reason.

\subsection{Data analysis}

Our study sessions yielded approximately 15 hours of audio that we transcribed. To analyze our interview and co-design session transcripts, we adopted a reflexive thematic analysis approach \cite{braun2019reflecting}. Two of the authors met after each interview and co-design session to conduct an interpretation session, and then conducted open coding of the transcripts. Throughout this coding process, the authors continuously discussed discrepancies in interpretation, and iteratively refined the codes based on these discussions \cite{braun2019reflecting,mcdonald2019reliability}.
In total, we generated around 1,125 unique codes. Through an iterative, bottom-up affinity diagramming process, we grouped codes into successively higher-level themes. The first level clustered our 1,125 codes into 271 themes. These were then clustered into 59 second-level themes, 15 third-level themes, and three final themes. We present our results in the following section, organized around our three final top-level themes.

\section{Findings}
\begin{figure*}[t]
  \centering
  \includegraphics[width=0.99\linewidth]{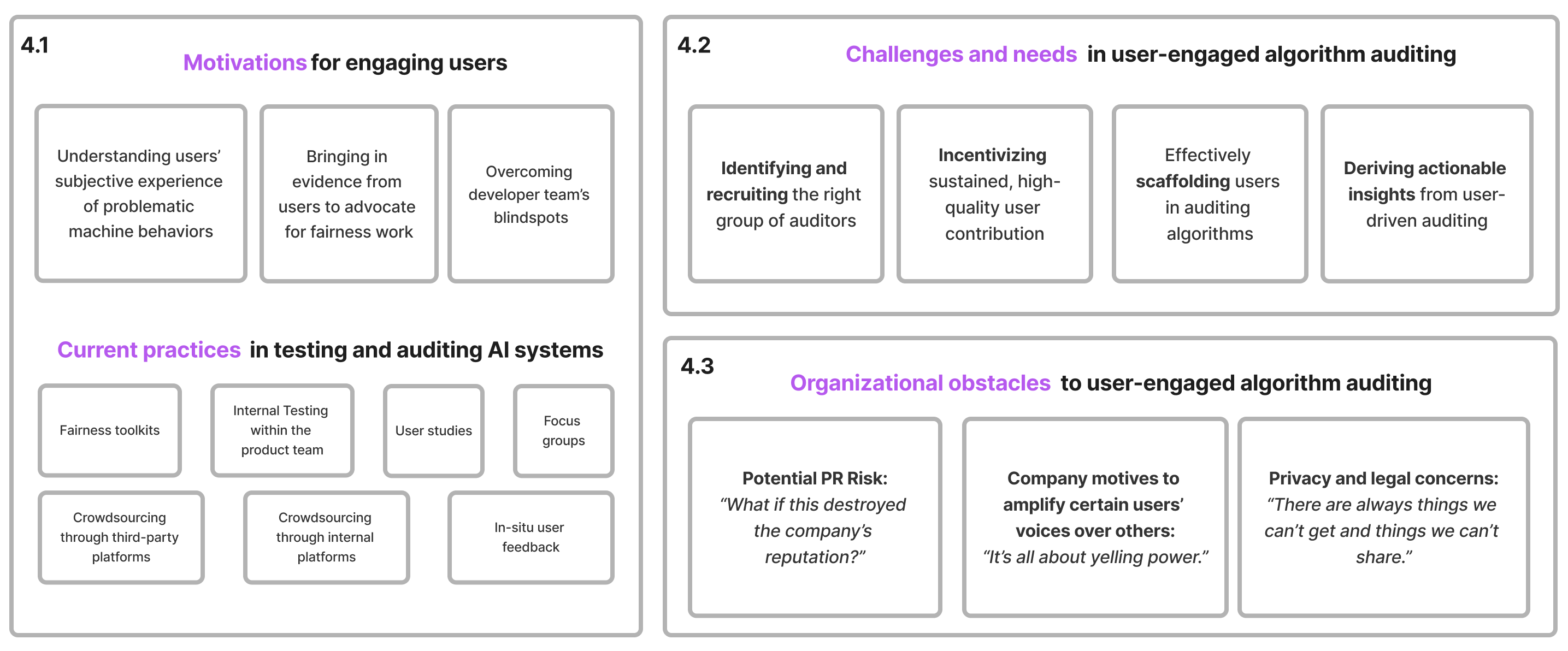}
  \caption{
    \rr{High-level overview of our findings. We first share participants’ existing motivations and practices for user engagement around algorithm testing and auditing (Section \ref{4.1}). We then describe challenges participants have faced in their attempts to effectively engage users (Section \ref{4.2}). Finally, we share broader organizational obstacles to user-engagement around algorithm auditing perceived by participants, highlighting key tensions that arise in practice when involving users in algorithm auditing efforts (Section \ref{4.3}).}
}
  \Description{An image showing the finding structure, containing participants’ existing motivations and practices for engaging users in auditing their AI products for harmful behaviors and biases, challenges participants faced in executing user-engaged auditing, and broader organizational obstacles to user-engaged auditing perceived by participants.}
  \label{fig:finding}
\end{figure*}

\rr{In this section, we describe how industry practitioners navigate the complicated process of engaging users in surfacing harmful algorithmic behaviors. As discussed below, practitioners mediate conflicts between (1) the underlying values of user engagement in algorithm testing and auditing, (2) inherent challenges in effective user engagement, and (3) the cultural, legal, and organizational obstacles that disincentivize bringing users' voices into algorithm auditing processes. We show that practitioners see clear advantages of engaging users in and around algorithm auditing processes, which have led them to explore leveraging crowdsourcing platforms and the design of new interfaces that enable in-situ feedback from their users. Yet practitioners also discussed complexities of engaging users in surfacing algorithmic harms, which introduce unique challenges beyond those faced with conventional human computation or user feedback systems. We summarize our findings in Figure \ref{fig:finding}.

}

\subsection{Practitioners’ motivations and practices for engaging users to audit AI products and services} \label{4.1}

\rr{Participants believed that engaging users in testing and auditing AI products and services could help them understand users’ subjective experiences of problematic machine behaviors and help to overcome developer teams’ blindspots. Participants also noted that having direct reports of potential issues from users could serve as powerful ammunition when making the case internally for a particular course of action. Driven by these motivations, participants reported having experimented with user engagement at various scales to test and audit their systems for problematic behaviors.}

\subsubsection{\textbf{Motivations to engage users in auditing AI products and services}} \label{4.1.1}

In our semi-structured interviews, all 12 participants reported that they currently use fairness toolkits or internal auditing with the developer teams to detect problematic behaviors in their AI products. However, participants quickly encountered limitations with these approaches, which prompted them to explore \rr{user-engaged} approaches as an alternative. Below, we discuss three major motivations that participants shared for experimenting with more \rr{user-engaged} approaches to algorithm auditing. 

\textbf{Understanding users’ subjective experiences of problematic machine behaviors: }
Participants were motivated to adopt \rr{user-engaged} approaches because in many settings, they felt \rr{that} it was not possible to measure ``unfairness’’ or ``harmfulness’’ without an understanding of users’ perceptions. As shown in Table \ref{tab:approaches}, eight of our 12 participants had previously used open-source ML fairness toolkits (e.g., Fairlearn \cite{FairlearnAPI}, AIF360 \cite{AIF360API}) to attempt to audit their teams’ AI systems. Participants said these toolkits only offered aggregate fairness metrics, whereas appropriately assessing algorithmic biases and harms in their systems required in-depth, qualitative assessments (e.g., for representational harms \cite{crawford2017trouble, holstein2019improving}). For example, P3 shared that their main role initially involved quantitatively assessing potential biases in NLP applications using fairness toolkits. However, P3’s team soon sought out user voices, because \emph{“things like representational harms were difficult to quantify and analyze using fairness toolkits, without understanding the reason behind why users with certain identit[ies] feel uncomfortable or offended.”} To this end, P3’s team conducted focus groups with users who encountered harmful biases in their NLP applications, in order to better understand users’ \textit{experiences} in their own words. 

Several participants described sometimes needing help from users to \rr{\emph{determine}} what ``fair’’ or ``non-harmful’’ behavior would look like in the first place. For example, P9 said a common approach when auditing a recommendation system is to assess \emph{``how diverse the recommendations on your platform are.”} However, different users often have conflicting perspectives regarding what it means to have appropriately ``diverse’’ recommendations. P9 believed the way to understand this was to \emph{``directly ask them[users] what they think.”} Similarly, P11 emphasized that in order to empower their team to design effective remediations, \emph{``It’s not enough to just have people flag that a recommendation is stereotypical, we want to understand why they think it is so that our team could brainstorm potential solutions.”}

\textbf{Overcoming developer teams’ blindspots:}
Another motivation for adopting \rr{user-engaged} testing and auditing approaches was to overcome cultural and experiential blindspots among product team members. Ten of our 12 participants said they conducted internal testing within the product team (see Figure \ref{tab:approaches}, column 2). However, echoing findings from Holstein et al. \cite{holstein2019improving}, participants reported that internal testing with small groups of developers often resulted in blindspots, prompting them to involve users to surface \emph{``unknown unknowns”} (P10), or the issues that the team does not know exist. For example, P8 noted that the images on their platform came from all over the world, involving signs and languages that were deeply cultural and regional, so \emph{``it would never be possible for [their] team to capture all these diverse aspects.”} Similarly, P5 complained that ``a lot of [their] AI ethics and bias activities only contain [their] own employees, and the perspectives are extremely limited.” P5 found that bringing in perspectives from external stakeholders helped to surface problematic algorithmic behaviors that their internal teams had never considered.

Some participants specifically emphasized the importance of engaging marginalized community members in auditing their AI products. P12 said, \emph{``No one in the developing team speaks the language and knows the idioms — how would they properly audit the outcomes? That’s why I have been spending time bringing native speakers into the auditing process.”} When P11’s team first started auditing the recommendation system in their product, they attempted to use personas to simulate real-world stakeholders and bring empathy to the team members, but they soon realized their blindspots persisted, as \emph{“the hypothetical personas are still basically what we imagined based on our understanding […] it’s just not realistic for a group of white people to truly understand Black artists’ perspectives through fictional cards.”}

\begin{table*}[]
\centering
\resizebox{\textwidth}{!}{%
\begin{tabular}{|l|
>{\columncolor[HTML]{D9D9D9}}c 
>{\columncolor[HTML]{D9D9D9}}c |
>{\columncolor[HTML]{D9EAD3}}c 
>{\columncolor[HTML]{D9EAD3}}c 
>{\columncolor[HTML]{B6D7A8}}c 
>{\columncolor[HTML]{B6D7A8}}c 
>{\columncolor[HTML]{B6D7A8}}c |}
\hline
 &
  \multicolumn{2}{l|}{\cellcolor[HTML]{D9D9D9}Approaches that do not engage users} &
  \multicolumn{5}{c|}{\cellcolor[HTML]{D9EAD3}\rr{User-engaged} approaches \rr{for testing and auditing their AI systems}} \\ \hline
 &
  \multicolumn{1}{l|}{\cellcolor[HTML]{D9D9D9}Fairness toolkits} &
  \multicolumn{1}{>{\centering\arraybackslash}p{2.5cm}|}{\cellcolor[HTML]{D9D9D9} Internal testing within the product team} &
  \multicolumn{1}{l|}{\cellcolor[HTML]{D9EAD3}User studies} &
  \multicolumn{1}{l|}{\cellcolor[HTML]{D9EAD3}Focus groups} &
  \multicolumn{1}{>{\centering\arraybackslash}p{2.5cm}|}{\cellcolor[HTML]{B6D7A8}Crowdsourcing through third-party platforms} &
  \multicolumn{1}{>{\centering\arraybackslash}p{2.5cm}|}{\cellcolor[HTML]{B6D7A8}Crowdsourcing through internal platforms} &
  In-situ user feedback \\ \hline
P1 &
  \multicolumn{1}{c|}{\cellcolor[HTML]{D9D9D9}X} &
  X &
  \multicolumn{1}{l|}{\cellcolor[HTML]{D9EAD3}} &
  \multicolumn{1}{c|}{\cellcolor[HTML]{D9EAD3}X} &
  \multicolumn{1}{l|}{\cellcolor[HTML]{B6D7A8}} &
  \multicolumn{1}{c|}{\cellcolor[HTML]{B6D7A8}X} &
  \multicolumn{1}{c|}{\cellcolor[HTML]{B6D7A8}X} \\ \hline
P2 &
  \multicolumn{1}{l|}{\cellcolor[HTML]{D9D9D9}} &
  X &
  \multicolumn{1}{c|}{\cellcolor[HTML]{D9EAD3}X} &
  \multicolumn{1}{l|}{\cellcolor[HTML]{D9EAD3}} &
  \multicolumn{1}{l|}{\cellcolor[HTML]{B6D7A8}} &
  \multicolumn{1}{l|}{\cellcolor[HTML]{B6D7A8}} &
   \\ \hline
P3 &
  \multicolumn{1}{c|}{\cellcolor[HTML]{D9D9D9}X} &
  X &
  \multicolumn{1}{l|}{\cellcolor[HTML]{D9EAD3}} &
  \multicolumn{1}{c|}{\cellcolor[HTML]{D9EAD3}X} &
  \multicolumn{1}{c|}{\cellcolor[HTML]{B6D7A8}X} &
  \multicolumn{1}{l|}{\cellcolor[HTML]{B6D7A8}} &
   \\ \hline
P4 &
  \multicolumn{1}{c|}{\cellcolor[HTML]{D9D9D9}X} &
  \multicolumn{1}{l|}{\cellcolor[HTML]{D9D9D9}} &
  \multicolumn{1}{c|}{\cellcolor[HTML]{D9EAD3}X} &
  \multicolumn{1}{c|}{\cellcolor[HTML]{D9EAD3}X} &
  \multicolumn{1}{l|}{\cellcolor[HTML]{B6D7A8}} &
  \multicolumn{1}{l|}{\cellcolor[HTML]{B6D7A8}} &
   \\ \hline
P5 &
  \multicolumn{1}{c|}{\cellcolor[HTML]{D9D9D9}X} &
  X &
  \multicolumn{1}{c|}{\cellcolor[HTML]{D9EAD3}X} &
  \multicolumn{1}{c|}{\cellcolor[HTML]{D9EAD3}X} &
  \multicolumn{1}{c|}{\cellcolor[HTML]{B6D7A8}X} &
  \multicolumn{1}{c|}{\cellcolor[HTML]{B6D7A8}X} &
   \\ \hline
P6 &
  \multicolumn{1}{l|}{\cellcolor[HTML]{D9D9D9}} &
  X &
  \multicolumn{1}{c|}{\cellcolor[HTML]{D9EAD3}X} &
  \multicolumn{1}{c|}{\cellcolor[HTML]{D9EAD3}X} &
  \multicolumn{1}{l|}{\cellcolor[HTML]{B6D7A8}} &
  \multicolumn{1}{l|}{\cellcolor[HTML]{B6D7A8}} &
   \\ \hline
P7 &
  \multicolumn{1}{l|}{\cellcolor[HTML]{D9D9D9}} &
  X &
  \multicolumn{1}{c|}{\cellcolor[HTML]{D9EAD3}X} &
  \multicolumn{1}{l|}{\cellcolor[HTML]{D9EAD3}} &
  \multicolumn{1}{c|}{\cellcolor[HTML]{B6D7A8}X} &
  \multicolumn{1}{l|}{\cellcolor[HTML]{B6D7A8}} &
  \multicolumn{1}{c|}{\cellcolor[HTML]{B6D7A8}X} \\ \hline
P8 &
  \multicolumn{1}{c|}{\cellcolor[HTML]{D9D9D9}X} &
  \multicolumn{1}{l|}{\cellcolor[HTML]{D9D9D9}} &
  \multicolumn{1}{l|}{\cellcolor[HTML]{D9EAD3}} &
  \multicolumn{1}{c|}{\cellcolor[HTML]{D9EAD3}X} &
  \multicolumn{1}{l|}{\cellcolor[HTML]{B6D7A8}} &
  \multicolumn{1}{c|}{\cellcolor[HTML]{B6D7A8}X} &
  \multicolumn{1}{c|}{\cellcolor[HTML]{B6D7A8}X} \\ \hline
P9 &
  \multicolumn{1}{c|}{\cellcolor[HTML]{D9D9D9}X} &
  X &
  \multicolumn{1}{c|}{\cellcolor[HTML]{D9EAD3}X} &
  \multicolumn{1}{c|}{\cellcolor[HTML]{D9EAD3}} &
  \multicolumn{1}{c|}{\cellcolor[HTML]{B6D7A8}} &
  \multicolumn{1}{c|}{\cellcolor[HTML]{B6D7A8}X} &
   \\ \hline
P10 &
  \multicolumn{1}{c|}{\cellcolor[HTML]{D9D9D9}X} &
  X &
  \multicolumn{1}{c|}{\cellcolor[HTML]{D9EAD3}X} &
  \multicolumn{1}{c|}{\cellcolor[HTML]{D9EAD3}X} &
  \multicolumn{1}{c|}{\cellcolor[HTML]{B6D7A8}} &
  \multicolumn{1}{c|}{\cellcolor[HTML]{B6D7A8}} &
  \multicolumn{1}{c|}{\cellcolor[HTML]{B6D7A8}X} \\ \hline
P11 &
  \multicolumn{1}{c|}{\cellcolor[HTML]{D9D9D9}X} &
  X &
  \multicolumn{1}{c|}{\cellcolor[HTML]{D9EAD3}X} &
  \multicolumn{1}{c|}{\cellcolor[HTML]{D9EAD3}} &
  \multicolumn{1}{c|}{\cellcolor[HTML]{B6D7A8}X} &
  \multicolumn{1}{c|}{\cellcolor[HTML]{B6D7A8}X} &
   \\ \hline
P12 &
  \multicolumn{1}{c|}{\cellcolor[HTML]{D9D9D9}X} &
  X &
  \multicolumn{1}{c|}{\cellcolor[HTML]{D9EAD3}} &
  \multicolumn{1}{c|}{\cellcolor[HTML]{D9EAD3}X} &
  \multicolumn{1}{c|}{\cellcolor[HTML]{B6D7A8}} &
  \multicolumn{1}{c|}{\cellcolor[HTML]{B6D7A8}X} &
   \\ \hline
\end{tabular}%
}
\Description[A table of 12 research participants’ existing approaches to auditing AI products and services.] {Column one lists the participant number. Columns two and three mark whether participants conducted auditing with an approach that does not engage users. Column two marks whether participants adopted fairness toolkits. The majority did. Column three marks whether the participant conducted internal testing within the product team. The majority did. Columns four to eight mark whether participants conducted \rr{user-engaged} auditing approaches. Column four marks whether participants adopted user studies. The majority did. Column five marks whether participants conducted studies among focus groups. The majority did. Column six marks whether participants conducted crowdsourcing through third-party platforms. Fewer than half did. Column seven marks whether participants conducted crowdsourcing through internal platforms. Half did. Column eight marks whether participants conducted In-situ user feedback. Fewer than half did.}
\caption{Participants' existing practices for \rr{testing and} auditing their AI systems for harmful behaviors and biases. In this table, we report both approaches that do not engage users and \rr{user-engaged} approaches. Section \ref{4.1.1} describes the \emph{limitations} of approaches that do not engage users. Section \ref{4.1.2} describes the \rr{user-engaged} approaches in detail. \rr{In this table, approaches that do not engage users are color-coded in gray; approaches for user-engagement at small scales are shown in light green; and approaches for user-engagement at larger scales are shown in darker green.}}
\label{tab:approaches}
\end{table*}

\textbf{Bringing in evidence from users to advocate for fairness work:}
Finally, five participants shared that one of their major motivations for adopting \rr{user-engaged} approaches was to gather evidence, in the form of direct quotes from users, that could help them persuade others on their teams that an issue was worth addressing. This was a particularly important motivation among three participants (P2, P6, P12) who were \textit{self-motivated} to address fairness issues in their teams’ products and not in roles that directly incentivized and supported this work. For example, P2 shared that, while building AI services to predict potential donors to their client institution, conversations with donors from marginalized communities inspired their team to \emph{“fundamentally reevaluate the potential biases towards who will donate in their dataset.”} As another example, P6 said their \emph{``engineers were pretty confident about the performance of [their] model and just making assumptions about the real-world situation when people are using the tool actively,”} rarely communicating with UX researchers like P6. However, P6 \emph{``got the luxury to chat directly with folks who are working on building the product”} when they brought in users’ voices and activities about the tool. Similarly, P4 shared that when data scientists voice concerns about the ML model being unfair and biased using numbers and graphs, \emph{``it is not a tangible risk to a business owner [...] bad experiences and feedback from the users work way better than numbers to motivate business owners to think deeper about the product’s potential negative impact.”}

\subsubsection{\textbf{Existing algorithm auditing approaches that engage with users}} \label{4.1.2}
Driven by the motivations discussed above, all 12 of our participants reported having experimented with \rr{user-engaged} approaches to \rr{test and} audit their systems for problematic algorithmic behaviors. As shown in Table \ref{tab:approaches}, these methods varied in scale: practitioners conducted user studies and focus groups with small groups of users, but they also attempted to engage a larger number of users in the auditing process through crowdsourcing tools or in-situ user feedback. 

\textbf{User studies and focus groups:} Nine participants shared that they have conducted single-person user studies, and eight said they have conducted focus groups, to engage users in testing their AI services and products. Participants shared how working with users provided them with different and new perspectives on a product and its potential biases. To expand their perspectives, some practitioners specifically sought out users who belonged to marginalized groups. For example, in order to mitigate potential biases in their donor database—which was used as the training data set for their AI service to identify potential future donors—P2’s team connected with prior donors who have minority backgrounds and were \emph{“not being included in the current database [that was] full of rich white people”} in order to understand their donating perspectives and experiences. This approach helped P2’s team realize that their database prioritizes privileged race and gender groups and change their overall problem formulation and development strategy for their models. Similarly, P12 helped their machine translation team better understand how errors in their products impacted marginalized immigrant communities through \emph{“chatting with immigrants living in the US with low language proficiency both individually and in groups.”}

\textbf{Crowdsourcing and in-situ user feedback approaches:} Although talking to users provided practitioners with new perspectives on potential harms their products might introduce, practitioners desired more scalable approaches to reach larger and more diverse groups. As a result, 9 of our 12 participants reported that they had previously leveraged \emph{crowdsourcing} or \emph{in-situ user feedback} approaches to attempt to engage users in surfacing harmful machine behaviors \emph{at scale}. This included using both third-party and internally-built crowdsourcing tools, as well as soliciting in-situ feedback directly from users during everyday interactions with their products and services. 

In our semi-structured interviews, some participants reported using third-party crowdsourcing platforms like Amazon’s Mechanical Turk or company-internal crowdsourcing platforms to invite a large number of crowd workers to help audit their AI products. For example, P3’s team found it useful to conduct focus group workshops to have users audit their NLP applications.  However, given the number of languages that their applications cover, their team turned to more scalable methods. For instance, to audit a model intended to detect offensive sentences, P3 deployed a task on Amazon’s Mechanical Turk to \emph{“ask crowd workers if they find the sentence offensive or not, and ask them to flag the offensive ones.”} This approach helped P3’s team uncover a large number of offensive sentences that were previously not predicted as offensive by their language models. Similarly, P5’s team paid crowdworkers on a third-party crowdsourcing platform to detect potential biases exhibited in their optical character recognition (OCR) model towards hand-written characters in different source languages. 

Inspired by recent cases in which users surfaced biases through their day-to-day interactions with the AI systems (e.g., \cite{shen2021everyday, chowdhury2021introducing}), four participants developed features that were \textit{seamlessly integrated into their AI systems’ user interfaces}, to allow users to provide feedback on potential biases and harmful behaviors in-situ, during their everyday interactions with a system. For example, P7 led an effort to build a plug-in to help users report any harmful outputs encountered while using their image recognition service. Similarly, P10 led an effort to build a web-based application where users could interact with an unreleased conversational agent prototype and report any problematic algorithmic behaviors they encountered during their interactions with the prototype. 

\subsection{Challenges and needs in \rr{user-engaged} algorithm auditing} \label{4.2}

\rr{Participants noted that although their current crowdsourcing and in-situ user feedback approaches could mitigate some of the limitations of existing auditing approaches, they still faced numerous challenges in effectively engaging users beyond existing design considerations for human computation or user feedback systems. For example, detecting harmful algorithmic behaviors requires recruiting and incentivizing the right group of auditors, with relevant cultural backgrounds, lived experiences, and perspectives. In addition, designing auditing tasks can introduce unique complexities, beyond those faced in conventional human computation tasks: the tasks need to guide users towards productive auditing strategies, but without overly influencing them to simply replicate industry practitioners' own biases and blind spots.}





\subsubsection{\textbf{Identifying and recruiting the right group of auditors}} \label{4.2.1}

All 12 participants emphasized the importance of  identifying and recruiting users with relevant identities, cultural backgrounds, and expertise to better test and audit the AI systems. 

\textbf{Identifying relevant subgroups of users:}
While discussing the design of the \emph{``developer-led”} \rr{user-engaged} auditing pipeline, many participants (N=6) shared that they found it challenging to determine which demographic subgroups were most critical to engage in auditing specific AI products. For example, P6’s team wanted to engage real-world users to audit their conversational AI products, yet they did not know \emph{“who are these users, how much they will be impacted by our product, and how to reach out to them.”} When P7’s team attempted to recruit users to assist in auditing their image search service, they became overwhelmed as they \emph{``started to think about the intersectionalities [of users], and the demographics just blew up into a billion different categories”} leading P7's team to wonder \emph{``what is the right level of identity intersections to look at?”} P7 noted that this challenge caused their team to \emph{“get lost”} in the process of setting up their own \rr{user-engaged} auditing pipeline. Similarly, P9, whose team worked on building fairer recommendation systems, shared, \emph{“We were having [a] hard time defining the genre of the content from the artists, and we don’t want to just label the artists by ourselves and project our biases [towards] the users even before we started to engage them.”} Given this experience, P9 viewed the challenge of identifying relevant subgroups of users as \emph{“a fundamental but intractable first step”} to conducting \rr{user-engaged} auditing on their product.

\textbf{Recruiting a diverse and representative set of users:} 
All 12 participants shared that, even when they knew what identities or background expertise they wanted to target, their current approaches were inadequate to actually recruit the targeted groups of users. Six participants reported that, in order to recruit user auditors, they relied heavily on personal networks or existing relationships with previous users. As such, they encountered challenges in recruiting users auditors from demographics, domains,  regions and cultures they had not established such relationships with. For example, P1 shared that, as a US-based company with few Asian employees, their team encountered challenges in recruiting \emph{``users from Japan and Korea”} to judge, for example, whether their AI services generate \emph{``potentially offensive labels for a sea area between Japan and Korea, given that these two countries had previous conflicts on the naming and jurisdiction of that area.’’} Similarly, P12 mainly relied on their team members’ existing personal relationships with customers to recruit marginalized community members, but often stumbled when attempting to reach out to a community that none of their team members had previously interacted with.

During the co-design sessions, several participants (N=9) expressed the belief that future tools that support more \emph{``user-led”} approaches to auditing, which appeal to users’ \textit{intrinsic} and \textit{social} motivations to participate, could be helpful in reaching users with specific identity characteristics and/or domain knowledge. For example, P5 drew an analogy to Wikipedia, believing that a more \emph{``user-led”} process could attract users to voluntarily and collaboratively audit AI systems that affect their lives, similar to how \emph{“people collectively edit articles on Wikipedia based on their interests, and will attract people with similar interests to join.”} Similarly, P12 saw potential for more \emph{“user-led”} approaches to \rr{user-engaged} algorithm auditing to organically attract people with similar identities and shared experience: \emph{“For example, people on Reddit with shared identities will come together and discuss problems they are facing.”}

\subsubsection{\textbf{Incentivizing sustained, high-quality user contributions}} \label{4.2.2}

While previous work has shown both external and intrinsic motivators driving online collective actions \cite{hossain2012users, ling2005using, kraut2011encouraging}, in our study, we found that practitioners currently rely primarily upon external motivators, such as financial and social incentives, to motivate user auditors to make sustained, high-quality contributions. Meanwhile, participants shared challenges in employing these motivators.

\textbf{Challenges in implementing financial incentivization in more ``user-led” pipeline: } All 12 of our participants shared that they currently provide financial compensation to motivate user auditors. While participants found it relatively straightforward to implement financial incentivization in a more ``developer-led” auditing pipeline, during the co-design activities, five participants raised challenges around how to compensate users in a ``user-led” auditing pipeline that was more exploratory and discussion-based. For example, P7 commented that the \emph{``open-endedness’’} in the ``user-led” auditing pipeline also made it difficult to decide how to compensate user auditors: \emph{``How do you pay people in this context? Right? Like what with the task-based things, it's like, there's a clear incentive: you do the tasks, you get paid. Here, where it's exploratory, [do] you pay people for just spending time in this interface? Do you pay them for just chatting? Or just generating hypotheses?’’} Similarly, P10 shared a prior experience where their team \emph{``had discussed paying the users who gave good amounts [of] and quality feedback’’}, but the conversation died when their team struggled to come up with clear definitions of ``good amounts’’ and ``quality’’ \rr{to implement a concrete compensation plan.} 

\textbf{Benefits and risks of social motivators: } P1’s team leveraged social motivators like ``peer recognition” and ``social interaction”  by creating a ``star system’’ to reward high quality auditors, as they found users enjoyed earning stars to demonstrate their ``high reputation.” P1’s team further \emph{``implemented a leader board to keep track of who is bringing in the best feedback, the most feedback, and whose feedback is being endorsed by lots of other auditors,’’} and to allow users to write \emph{``recommendation letters”} for one another to audit other tasks. During the co-design study, several participants (N=8) believed that a more ``user-led” auditing pipeline could potentially amplify social motivators. Nevertheless, participants raised concerns that \emph{``certain user groups’ voices might be further marginalized in [the `user-led'] pipeline’’} (P9). To combat this, participants suggested developers should intervene and facilitate the conversations among the user auditors, to amplify marginalized voices throughout \rr{user-engaged} auditing processes. 

\subsubsection{\textbf{Effectively scaffolding users in auditing algorithms}} \label{4.2.3}
Participants noted that it was challenging, in practice, to design user auditing tasks and instructions that could empower users to generate meaningful insights about their AI products and services. As we discuss below, participants shared several challenges they had faced in guiding user auditors without imposing the development team’s own biases upon them, and in prompting user auditors to provide more critical feedback on an AI system’s overall design. 

\textbf{Guiding users towards productive auditing strategies, without overly biasing them:}
During the interview portion of our study, participants shared experiences where user auditors had misunderstood the tasks they were given or had failed to provide sufficient detail and context for industry teams to act on their reports. Thus, throughout the co-design portion of our study, several participants (N=8) expressed desires for ways to help user auditors better understand their team’s intended goals for an audit, and to scaffold them in auditing a system more effectively. For instance, P5 suggested \emph{``sharing with the users a theoretical structure of biases, an algorithmic harm taxonomy”} to reference both during an onboarding phase and at any point during their auditing activities. However, P5 was uncertain what such a taxonomy would look like in their context (i.e., sentiment analysis). Similarly, P8 and P10 desired better ways to nudge user auditors to \emph{``think out of the box’’} (P8) and \emph{``ask hard questions [to a conversational agent] and break the model and surface our `unknown unknown'”} (P10). P7 noted that, in the context of image search, user auditors may not always test the impacts of small perturbations, so they suggested prompting users to do so: \emph{``Chang[ing] a small word might lead to a very different search result, and we definitely want to guide users to explore these small changes’} (P7).

Despite this desire to guide user auditors in more productive directions, some participants (N=6) expressed concerns that providing too much or the wrong kinds of scaffolding might bias users to think too much like their own teams—potentially limiting the value of a \rr{user-engaged} auditing approach. For example, during our co-design activity, P10 noted that designing guidelines and specific prompts for user auditors was \emph{``quite tricky since of course we want to offer detailed guidelines and ask specific questions like `do you think this output is biased towards Asians or women,” but our questions might actually bias the users when they are finding biases. [...] We need to make sure that we don’t let our confirmation bias affect this [\rr{user-engaged} audit process]”}. 

Acknowledging the challenge of navigating these tradeoffs, five of our participants emphasized the importance of cross-functional collaboration in designing effective guidelines, tasks, and prompts for user auditors. For example, in most of these teams, the design of auditing tasks was left to engineers who had no training in the design of human subjects research methods. By contrast, P10 worked on a team where UX practitioners were involved in the design of auditing tasks. However, the UX and AI teams often worked in silos. P10 felt that \emph{``with a UX background, I only have a surface understanding of large language model[s]. I can’t design a good auditing task just by myself if we eventually want to incorporate some auditing feedback from users into the current model.”}

\textbf{Soliciting critical and holistic feedback from user auditors: }Ten of our 12 participants expressed desires for better ways to prompt more critical and holistic feedback from user auditors. While iterating on the \rr{user-engaged} audit report and the ``developer-led” auditing pipeline, P9 suggested that in order to prompt critical feedback, it could be helpful to share the team’s \textit{rationales} behind particular design decisions, which might otherwise remain opaque to user auditors. P9 shared a prior experience where their team had initially struggled to gather the sorts of critical feedback they were hoping for. However, after their team shared more details about specific design rationales, \emph{``[user auditors] asked questions like, ‘Do I actually want to get recommendations in this way? What’s a better way to design this that fits my preference?’ These types of questions are the ones we wish our users [would] ask when auditing and include [in] their final report.’’} Similarly, while describing their team’s prior experiences with \rr{user-engaged} audits, P7 mentioned that users sometimes express a desire to know more about \emph{``why and how’’} their AI products and services are designed, in addition to seeing the AI’s outputs, so that they could surface potential \textit{procedural} issues that might not otherwise be as visible to them.

Several participants believed that a more ``user-led” algorithm auditing process could help to catalyze more critical inquiries from users. For example, while iterating on the ``user-led’’ audit pipeline, P12 said, \emph{``when it’s `developer-led,’ we are still testing if there are mistakes or unfairness in the product, right? But what if users believe this AI product shouldn’t even exist? This is something you can get by giving people more freedom to discuss.’’} For this reason, P1 believed it was critical to \emph{``allow [user auditors] to chat with each other,’’} in contrast to conventional crowdsourcing approaches, in which crowdworkers perform tasks in independent silos. Similarly, P10 argued that platforms or tools for \rr{user-engaged} algorithm auditing should include mechanisms \textit{``for users to share and discuss the issue they found during the auditing process’’} 

\subsubsection{\textbf{Deriving actionable insights from \rr{user-engaged} auditing}} \label{4.2.4}

Seven out of 12 participants shared challenges they currently face in deriving actionable insights from the user audit reports. In particular, participants shared that, unlike in conventional crowdsourcing approaches, understanding the perspectives of ``outliers” may often be more important than understanding the majority view. In addition, participants found challenges in communicating qualitative auditing results to key decision-makers, given an organizational culture of valuing numbers over more complex stories.

\textbf{‘It is no longer simply checking the majority vote.” Aggregating and interpreting \rr{user-engaged} auditing reports: }Throughout our co-design activities, all 12 participants highlighted the challenges of aggregating and interpreting results from \rr{user-engaged} auditing processes. As P3 put it, \emph{``in more traditional numerical crowdsourcing activities, you would throw away a person who like always contradicts what everyone else says.’’} However, P3 noted that the ``outliers’’ in a \rr{user-engaged} audit are often the ones that developer teams care the most about. These ``outliers’’ may represent users in the margins, who are sensitized to issues that other auditors are not: \emph{``So maybe like everyone said, sample A was not offensive or problematic, except for like, auditor number 39 [...] because number 39 actually found problematic things others didn't’’} (P3). Similarly, P2 said that in their view, \emph{``a few [user] audit results stating potential biases and harms might weigh more than one hundred similar good audit results”}, similar to doing UX tests when \emph{``a single negative review might surface key insights for room to improve.”} 

The challenge of interpreting results in aggregate becomes especially hard when the number of user audit reports gets large. P7 stressed the importance of collecting \textit{```why’ information’’} from users, such as open-text responses explaining why a user perceives a particular algorithmic behavior as problematic. However, P7 complained that they currently lack an \emph{``efficient mechanism to combine quantitative and qualitative insights from users’ feedback [...] it is no longer simply checking the majority vote.”} While discussing the \rr{user-engaged} audit report, P5 said, \emph{``When I have thousands of training data annotations from crowd workers, I could just check the statistics of the aggregated results [...] but now with thousands of these audit reports, how am I supposed to figure out the most valuable information?”} P5 believed that in order to institutionalize \rr{user-engaged} auditing in their organization, their team needed to invest in developing new automatic pipelines to augment their current manual process of reviewing \rr{user-engaged} audit reports.

\textbf{``It’s just our current culture, we still believe more in the numbers’’: Quantification and its challenges: }All 12 of our participants mentioned that, in order to effectively integrate findings from a \rr{user-engaged} audit, they need to be able to present clear, quantifiable metrics to leadership and other team members. As P11 put it, \emph{``It’s just our current culture; we still believe more in the numbers.’’} However, this often presented challenges in the context of \rr{user-engaged} algorithm auditing. For example, when P12 could not offer a clean cut number but only \emph{``a complex story”} in response to questions from developers such as \emph{``[what] percentage of users believe their recommendation is bad,”} they were told by their product manager that they were being \emph{``distracting and counterproductive”} to project progress. While iterating on the \rr{user-engaged} auditing report, P6 shared related concerns about how to measure the progress of a \rr{user-engaged} audit: \emph{``One consideration for combining these \rr{user-engaged} reports is, what is the metric to define `success’ here? How many [reports] is enough? How much more do we need before we stop?”} P6 suggested that in order to effectively translate these reports into concrete actions from the product team, defining such metrics and clearly scoping the goal of a \rr{user-engaged} audit would be critical given that \emph{``[monetary and time] cost is [always a] concern."}

\subsection{\textbf{Organizational obstacles to \rr{user-engaged} algorithm auditing}} \label{4.3}

\rr{Beyond challenges in effectively engaging users, participants also shared broader, organizational obstacles they perceived around potential PR risks, profit motives that work against protecting marginalized groups, and privacy and legal concerns. Taken together, our findings shed light on the ways practitioners currently navigate organizational tensions specifically around user-engagement in algorithm testing and auditing.}

\subsubsection{\textbf{``What if this destroyed the company’s reputation?’’ Potential PR risk}} \label{4.3.1}

Multiple participants (N=6) raised concerns regarding the feasibility of full institutional buy-in to \rr{user-engaged} auditing from their organizations. Participants were especially skeptical of the more ``user-led'' pipeline, as it seemed to hold the greatest potential PR risk. For example, while co-designing the ``user-led'' pipeline, P7 shared, \emph{``PR issue[s are…] one main reason I see companies don’t want [user-led algorithm auditing] as an everyday thing.”} P7 backed this fear with a specific experience: \emph{``We had users just try to find the absolute worst thing possible in our models and [they] made it into a story for social media instead of reporting back to us.”} P7 also said their team leadership worried that this behavior could \emph{``expose the vulnerability of [their] models to their competing companies”} and that ultimately \emph{“involving users might create more headlines that damage the company’s public image”}. Similarly, P3 said these fears constituted \emph{``a major reason why [their] company is still experimenting [with \rr{user-engaged} approaches] on some applications instead of making it a company-wide thing.”} Even P8, whose company began \rr{user-engaged} auditing after PR pressure, worried that \rr{user-engaged} auditing could cause new PR issues. After co-designing the report, P8 said,\emph{``This report would be extremely useful if it only goes to us,”} and wondered how to \emph{“hold the users accountable and make sure they don’t destroy the company’s reputation after gaining the trust.”} 

\subsubsection{\textbf{``It’s all about yelling power’’: Company motives to amplify certain users’ voices over others}} \label{4.3.2}

An important goal of \rr{user-engaged} algorithm auditing is to translate the problems users find into concrete remediations from the product teams. However, eight of our 12 participants mentioned that, realistically, companies will prioritize addressing issues raised by certain groups of users over others. For instance, P1 said,\emph{``If [...] a large group of researchers from [a major US-based research institute] and a single user from a community college both raised concerns about our knowledge graph, unfortunately our business team would have to prioritize the former.”} P1 concluded that when addressing issues raised in \rr{user-engaged} audits, \emph{``It’s all about [the] yelling power of the users.”} During the co-design activity, P12 stated that in order to incorporate \rr{user-engaged} audit in their day-to-day AI work, \emph{``the biggest challenge is not to design the perfect workflow, but to make sure [the company] wants to do [\rr{user-engaged} auditing] for social good, [not just] for earning more money from more people.”} Similarly, P4 shared that in their current \rr{user-engaged} auditing work, they constantly find themselves battling the business teams since \emph{``the business teams sometimes choose to neglect the users’ audit outcome if the reports were not from their `original [target] audience.’”} P4 believed that, if their team is not implementing remediation in response to reports from the most marginalized users, their organizations \emph{``run the risk of participating in an ethics-washing activity.”}

Several participants also shared that they often ran up against a ``vicious cycle’’ (P12), in which a dearth of data from low-resource areas and marginalized communities makes it difficult for practitioners to advocate for more resources to address these areas. P12 said they struggled to get enough resources to test their language technologies with marginalized communities, as their data scientists required large-scale, quantitative evidence before approving studies with new groups of community members. Yet as P12 noted, \emph{``There is, of course, not enough evidence, when these groups were not even considered as users in the first place.’’} 

\subsubsection{\textbf{``There are always things we can’t get and things we can’t share’’: Privacy and legal concerns}} \label{4.3.3}

As mentioned in Section 4.2.1, participants highlighted that access to user auditors’ demographics and other background information is critical in order to assign tasks to appropriate user auditors and to understand which perspectives are represented in reports from user auditors. However, all 12 participants also shared challenges around obtaining certain demographic information due to privacy and legal concerns—mirroring challenges that practitioners face in AI auditing work more broadly \cite{chen2019fairness, holstein2019improving, veale2017fairer}. However, beyond standard concerns around the collection and use of sensitive data for AI auditing, participants also shared concerns about the data user auditors might share on an auditing platform, as well as the data they might need to share with user auditors to enable effective audits. For example, when discussing the \emph{``user-led”} auditing pipeline, P7 noted the challenges that could arise if a user auditor were to \emph{``[take] somebody else's photo and share it with other users for the purpose of auditing''} or \emph{``[share] any users’ race and gender with other users if they don’t want to share.”} P10, on the other hand, brought up their internal concerns around \emph{``losing the competition [with other companies] on building large language models,’’} by exposing too many model details to user auditors. As P10 shared at the end of the co-design session, \emph{``There are always things we can’t get and things we can’t share.”}

\section{Limitations}
Similar to prior HCI work studying responsible AI practices in industry (e.g., \cite{holstein2019improving,madaio2020co,rakova2021responsible}), our findings shed light on current practices, challenges, and needs among a set of practitioners who may be at the \textit{forefront} of an emerging industry practice. As discussed in Section \ref{Methods}, we recruited participants using a purposive sampling approach \cite{campbell2020purposive}. All of our participants were passionate about addressing harmful behaviors in their AI systems, and they had direct experience experimenting with \rr{user-engaged} approaches to algorithm auditing in their work. 
Furthermore, most of our participants worked at large technology companies, and all of our participants were located in the US (see Table \ref{tab:participants}). 


\section{Discussion}
Drawing upon prior literature in areas such as algorithm auditing, crowdsourcing, participatory design, and fairness in AI, we discuss opportunities for future HCI research to help realize the potential and mitigate the risks of \rr{user-engaged} auditing in industry practice.

\subsection{\rr{Unique challenges in supporting user-engaged algorithm auditing}} \label{6.2}

Prior research demonstrates great potential for human computation and crowdsourcing approaches to advance AI research and practice, especially in areas such as data generation and annotation \cite{russakovsky2015imagenet, callison2010creating} and human-level evaluation \cite{vaughan2017making, anastasiou2011comparison, zaidan2011crowdsourcing, zhou2019hype}. In our study, we found that industry practitioners’ current approaches to \rr{user engagement in} algorithm auditing are often built atop existing crowdsourcing pipelines (Section \ref{4.1.2}). However, practitioners quickly ran up against limitations of conventional \rr{human computation and crowdsourcing} approaches (Section \ref{4.2}). Below, we highlight \rr{five} design \rr{implications} for \rr{user-engaged} auditing that \rr{extend beyond standard considerations for human computation} approaches, and discuss corresponding directions for future HCI research.

\subsubsection{\textbf{\rr{Focusing on ``who''}}}
First, crowdsourcing approaches used to support AI research and practice typically focus more on \textit{what crowd workers do} and less on \textit{who they are} \cite{vaughan2017making, russakovsky2015imagenet, anastasiou2011comparison, callison2010creating, snow2008cheap}. However, the \emph{``who”} factors that encompass users’ intersectional identities and lived experience play critical roles in \rr{user-engaged} auditing, as discussed in Sections \ref{4.1.1} and \ref{4.2.1}. Prior work has also shown that users’ personal experiences with and exposures to bias influence the ways they search for and make sense of harmful behaviors in algorithmic systems \cite{DeVos2022TowardUA, eslami2015i, eslami2016first, buolamwini2018gender}. Future research should explore better processes and tools to support practitioners in identifying and recruiting appropriately diverse and representative user auditors, for particular algorithm auditing tasks.

\subsubsection{\textbf{\rr{Supporting sustained, long-term contributions}}}

Second, in contrast to the relatively transient, on-demand nature of most crowd work \cite{Kittur2013TheFO, bigham2015human}, \rr{user-engaged} auditing often requires sustained, long-term contributions from user auditors, to continuously improve AI systems (Section \ref{4.2.3}). This requirement entails radically different designs on both the process and interface levels. For example, future research may explore the interaction design space of in-situ feedback mechanisms, to solicit user auditors’ feedback in the context of their day-to-day interactions with algorithmic systems. A key challenge for this line of exploration is to solicit feedback on algorithmic behavior in formats that are quick and relatively unobtrusive to collect from users, yet at the same time are readily interpretable and actionable for AI practitioners on the receiving end \cite{groce2013you, attenberg2015beat}. \rr{To better support long-term contributions in user-engaged auditing, it is also critical to design intuitive and efficient interactive interface for users auditors \cite{wu2019errudite}}. In addition, future research may explore the design of social platforms to build sustained online communities of users who are motivated to engage in testing and auditing algorithmic systems together \cite{shen2021everyday}. 

\subsubsection{\textbf{\rr{Navigating inherent ambiguities in auditing tasks and outcomes}}}
Third, traditional crowdsourcing typically starts with well-defined goals and anticipated outcomes set up by the ``requesters’’ \cite{kittur2013future}. In contrast, there may be benefits to empowering users to collectively take the lead in directing auditing efforts (Section \ref{4.2.3}). With too much direction and guidance from industry AI teams, \rr{user-engaged} audits risk replicating the very biases and blindspots that they were meant to overcome (Section \ref{4.2.3}). \rr{Prior HCI research has mainly focused on developing tools and processes to better prompt users discovering more ``unknown unknowns" \cite{attenberg2011beat, suh2019anchorviz, wu2019errudite}}. Future research should explore the design of scaffolding mechanisms for user auditors that can navigate the trade-offs between promoting more effective algorithm auditing behaviors versus providing too much direction, limiting the kinds of issues user auditors are able to surface.

\subsubsection{\textbf{\rr{Reconsidering aggregation and quantification approaches to ensure marginalized voices are heard}}}
In addition, as discussed in recent HCI research (e.g., \cite{gordon2022jury}), typical crowdsourcing approaches involve aggregate analyses and evaluations, for example, by relying on ``majority vote’’ from the crowd in order to arrive at the results. Practitioners emphasized that relying exclusively on quantification to derive actionable insights from \rr{user-engaged} auditing could be harmful and counterproductive (Section \ref{4.2.4}). Resonating with findings from prior research \cite{madaio2021assessing, deng2022exploring}, we found that practitioners desired practical tools to support them in advocating for marginalized communities through an integration of both qualitative and quantitative forms of evidence (Section \ref{4.3.2}). Strategies like ``tactical quantification’’ proposed by Irani et al. \cite{irani2013turkopticon} could support practitioners in advocating on behalf of user auditors within their organizations, in industry contexts where numbers are culturally valued over more complicated stories. Future HCI research should explore the design of new tools, computational techniques, and visualization approaches that can aid practitioners in persuasively advocating on behalf of marginalized groups of users (cf. \cite{gordon2022jury,robertson2020-what,shen2022model}). 

\subsubsection{\textbf{Designing user-engaged auditing mechanisms with teeth}}

\rr{
Finally, prior work has emphasized the activist nature of auditing, which differs from much prior research on human computation and crowdsourcing \cite{metaxa2021auditing}. The end goal for user-engaged auditing is to improve products and protect future users from algorithmic harms. Yet achieving this goal requires that companies are actually held accountable for addressing the issues that user auditors uncover \cite{buolamwini2018gender, Tay_Ban, Facebook_Ban, raji2020closing}. Thus, it is critical for future HCI research to design for user-engaged auditing with accountability and collective empowerment in mind. For example, researchers could explore building platforms that support user auditors in collectively applying pressure and holding companies to account, when serious issues are not addressed \cite{salehi2015we, irani2013turkopticon}. In the next section, we further expand on the discussion around these design implications. 
}

\subsection{The complex relationship between industry practitioners and user auditors} \label{6.1}

Our study explored industry practitioners’ perceptions of and relationships with \rr{user-engaged} approaches to algorithm auditing. A key component of this is the relationship between practitioners and the user auditors who power the auditing process. As these two groups strive to surface and address harmful algorithmic behaviors, a complex relationship and a delicate \emph{mutual (dis)trust} is revealed that can be friendly or antagonistic, depending on the situation.

\subsubsection{\textbf{How might users trust AI practitioners?}}
For user auditors, auditing is often seen as a form of activism, in which rooting out harmful behaviors in algorithmic systems benefits society \cite{metaxa2021auditing}. Past research on \rr{user-engaged} algorithm auditing highlighted users’ advocacy for marginalized groups of people, expressing solidarity via their auditing activities \cite{DeVos2022TowardUA}. However, industry practitioners in our study rarely brought up similar rationales for engagement with \rr{user-engaged} auditing (Section \ref{4.2.2}). This might be because industry practitioners have greater opportunity to directly effect change on issues in the algorithmic systems they work on, whereas users typically need to rely on their collective power to raise awareness in order to be heard \cite{shen2021everyday, metaxa2021auditing}. Though \rr{user-engaged} audits can empower users through the ability to directly connect with industry practitioners to try and identify issues together, they also firmly place the choice to take action with practitioners, potentially leaving users with less room for leverage via other means. 

Given this asymmetric power dynamic between user auditors and AI practitioners, how could HCI researchers support empowering users’ collective action when users’ needs are not met and their trust fractured? Previous platforms like Turkopticon \cite{irani2013turkopticon} and WeAreDynamo \cite{salehi2015we} demonstrated the potential for HCI researchers to consciously build spaces for activism, leveraging their collectiveness to negotiate their desires and needs. Incorporating similar spaces into future \rr{user-engaged} auditing processes could alleviate users’ concerns and empower them to act, ensuring that issues they collectively surface will be addressed in satisfactory ways.

\subsubsection{\textbf{How might AI practitioners trust users?}}
Despite the desire for holistic and critical auditing processes (Section \ref{4.2.3}), industry practitioners in our study expressed trust concerns around opening their systems to be audited by users (Section \ref{4.3.1}). As described above, users frequently turn to public awareness raising around problematic algorithmic behaviors they encounter as leverage to pressure companies into addressing those issues. How can practitioners fully trust users to audit their systems without publicizing issues before practitioners have had a chance to address them? Indeed, in our study, practitioners frequently cited apprehensions that user auditors might harm company reputations through negative PR (Section \ref{4.3.1}). However, other practitioners viewed themselves as more on the side of users. These practitioners work beyond their main job functions to advocate for users, while simultaneously using the issues brought by user auditors as evidence to convince their teams to act, putting pressure on companies from the inside. \rr{How might we support and protect practitioners who genuinely strive to mitigate or avoid problematic algorithmic behaviors in their AI products and services? To this end, }similar to supporting user activism, there may be opportunities to build platforms (cf.\cite{irani2013turkopticon, salehi2015we}) to support collective actions amongst these practitioner individual advocates within companies.

One avenue of exploration to address the tensions of required mutual trust might take inspiration from security bug bounties \cite{malladi2019bug}. In these, security experts turn over information about security vulnerabilities to companies in exchange for monetary compensation and the promise of a fix, with the understanding between parties that if the issue is not addressed in a given time frame, then the security expert will publicize the issue. Borrowing from this model could enhance the trust between user auditors and industry practitioners as well, allowing practitioners a protected timeframe to fix issues but enacting a strict deadline for users to take further action if practitioners’ promises are not upheld. Indeed, emergent projects like bias bounties \cite{chowdhury2021introducing, raji2022outsider}have begun to transfer some bug bounty success to the territory of \rr{user-engaged} algorithm auditing.

\rr{\subsection{Users' perspectives on user-engaged algorithm auditing} \label{6.3}

As prior research has begun to explore users’ practices and perspectives around user-engaged algorithm auditing \cite{DeVos2022TowardUA, shen2021everyday, lam2022enduser}, our study begins to fill the gap between this emerging literature and industry practitioners’ perspectives. However, while supporting practitioners in designing and implementing more effective forms of user-engagement in algorithm auditing, it is critical to continue to explore users' perspectives and values. To this end, future work should bring together industry practitioners' and users' perspectives, potentially through the collaborative design, development, and oversight of user-engaged auditing procedures and platforms. 

Importantly, when users are engaged in algorithm testing and auditing, they necessarily observe and likely experience some of those harms themselves. In our study, practitioners described thinking about better ways to target user auditors with relevant identity characteristics. This is in line with prior research, which has found that people with certain exposures and experiences are more able to surface related issues in algorithmic systems \cite{DeVos2022TowardUA}. These people may be ideal candidates to serve as user auditors. At the same time, since these are often members of marginalized communities, who are already overburdened and more likely to be the targets of harmful algorithmic behavior \cite{harrington2019deconstructing, sloane2020participation, hsu2022empowering}, they are also more likely to be harmed through participation in algorithm testing and auditing. Furthermore, drawing an analogy to platform content moderation, in which moderators are often exposed to violence and harassment and could be subject to long-term psychological harms \cite{steiger2021psychological, dosono2019moderation}, auditing for problematic algorithmic behaviors may also result in long-term psychological harms towards user auditors.

Therefore, future research should consider and design to alleviate potential emotional burdens and psychological harms toward user auditors. Furthermore, \rr{user-engaged} algorithm auditing could harm users if their labors are co-opted in ways that are not aligned with what they might have wanted. We highlight these burdens on users as vital areas for further research. Despite the burdens, \rr{user-engaged} auditing, when implemented well, can serve to reduce algorithmic harms present and acting in the world now. We urge continual evaluation of \rr{user-engaged} auditing processes by practitioners to ensure that these burdens and potential harms are mitigated. Future research should also explore the potential of computational or other alternative solutions to reduce the need for continuous involvement of the most vulnerable populations in the auditing process (e.g., \cite{gordon2022jury, lam2022enduser}), with the caution that computational approaches may also risk introducing new types of harms to users.
}


\section{Conclusion}
We conducted a series of interviews and iterative co-design activities with industry practitioners to gain insights into the current landscape and future opportunities for \rr{user-engaged} algorithm auditing in industry practice. We surfaced major motivations for engaging users in AI testing and auditing and described practitioners’ existing approaches for \rr{user-engaged} auditing. We found that practitioners face challenges around appropriately recruiting and incentivizing user auditors, scaffolding user audits, and deriving actionable insights from audit reports. Furthermore, practitioners shared broader organizational obstacles to \rr{user-engaged} auditing, highlighting key tensions that arise in practice when involving users in algorithm auditing efforts. Based on these findings, we discussed the complex relationships between practitioners and user auditors, offering potential remediation for developing mutual trust. We then describe various opportunities to support \rr{user-engaged} auditing beyond existing design considerations for human computation or user feedback systems. Overall, we hope that this work inspires future efforts to realize the potential and mitigate the risks of \rr{user-engaged} auditing in industry practice.



\begin{acks}
This work was supported by the National Science Foundation (NSF)
program on Fairness in AI in collaboration with Amazon under
Award No. IIS-2040942, an award from Cisco Research, and an award from the Jacobs Foundation. We would like to thank Alex Cabrera, Charvi Rastogi, Tzu-sheng Kuo, Kimi Wenzel, Seyun Kim, Katelyn Morrison, Adam Perer, Jason Hong for their feedback on the draft. Special thanks to our anonymous reviewers and to all participating industry practitioners for making this work possible.
\end{acks}

\balance

\bibliographystyle{ACM-Reference-Format}
\bibliography{citation, CHI22}


\begin{thebibliography}{95}


\ifx \showCODEN    \undefined \def \showCODEN     #1{\unskip}     \fi
\ifx \showDOI      \undefined \def \showDOI       #1{#1}\fi
\ifx \showISBNx    \undefined \def \showISBNx     #1{\unskip}     \fi
\ifx \showISBNxiii \undefined \def \showISBNxiii  #1{\unskip}     \fi
\ifx \showISSN     \undefined \def \showISSN      #1{\unskip}     \fi
\ifx \showLCCN     \undefined \def \showLCCN      #1{\unskip}     \fi
\ifx \shownote     \undefined \def \shownote      #1{#1}          \fi
\ifx \showarticletitle \undefined \def \showarticletitle #1{#1}   \fi
\ifx \showURL      \undefined \def \showURL       {\relax}        \fi
\providecommand\bibfield[2]{#2}
\providecommand\bibinfo[2]{#2}
\providecommand\natexlab[1]{#1}
\providecommand\showeprint[2][]{arXiv:#2}

\bibitem[Abebe et~al\mbox{.}(2020)]%
        {abebe2020roles}
\bibfield{author}{\bibinfo{person}{Rediet Abebe}, \bibinfo{person}{Solon
  Barocas}, \bibinfo{person}{Jon Kleinberg}, \bibinfo{person}{Karen Levy},
  \bibinfo{person}{Manish Raghavan}, {and} \bibinfo{person}{David~G Robinson}.}
  \bibinfo{year}{2020}\natexlab{}.
\newblock \showarticletitle{Roles for computing in social change}. In
  \bibinfo{booktitle}{\emph{Proceedings of the 2020 Conference on Fairness,
  Accountability, and Transparency}}. \bibinfo{pages}{252--260}.
\newblock


\bibitem[AI(2021a)]%
        {AIF360API}
\bibfield{author}{\bibinfo{person}{IBM Resaerch~Trusted AI}.}
  \bibinfo{year}{2021}\natexlab{a}.
\newblock \showarticletitle{AIF360 API}.
\newblock  (\bibinfo{year}{2021}).
\newblock
\urldef\tempurl%
\url{https://aif360.mybluemix.net/}
\showURL{%
\tempurl}


\bibitem[AI(2021b)]%
        {AIX360API}
\bibfield{author}{\bibinfo{person}{IBM Resaerch~Trusted AI}.}
  \bibinfo{year}{2021}\natexlab{b}.
\newblock \showarticletitle{AIX360 API}.
\newblock  (\bibinfo{year}{2021}).
\newblock
\urldef\tempurl%
\url{https://aix360.mybluemix.net/}
\showURL{%
\tempurl}


\bibitem[AI(2022)]%
        {ChatGPT_Feedback}
\bibfield{author}{\bibinfo{person}{Open AI}.} \bibinfo{year}{2022}\natexlab{}.
\newblock \bibinfo{title}{ChatGPT Feedback Contest: Official Rules}.
\newblock
\newblock
\urldef\tempurl%
\url{https://cdn.openai.com/chatgpt/ChatGPT_Feedback_Contest_Rules.pdf}
\showURL{%
\tempurl}


\bibitem[Anastasiou and Gupta(2011)]%
        {anastasiou2011comparison}
\bibfield{author}{\bibinfo{person}{Dimitra Anastasiou} {and}
  \bibinfo{person}{Rajat Gupta}.} \bibinfo{year}{2011}\natexlab{}.
\newblock \showarticletitle{Comparison of crowdsourcing translation with
  Machine Translation}.
\newblock \bibinfo{journal}{\emph{Journal of Information Science}}
  \bibinfo{volume}{37}, \bibinfo{number}{6} (\bibinfo{year}{2011}),
  \bibinfo{pages}{637--659}.
\newblock


\bibitem[Angwin et~al\mbox{.}(2016)]%
        {angwin2016machine}
\bibfield{author}{\bibinfo{person}{Julia Angwin}, \bibinfo{person}{Jeff
  Larson}, \bibinfo{person}{Surya Mattu}, {and} \bibinfo{person}{Lauren
  Kirchner}.} \bibinfo{year}{2016}\natexlab{}.
\newblock \showarticletitle{Machine bias}.
\newblock \bibinfo{journal}{\emph{ProPublica}} (\bibinfo{date}{May}
  \bibinfo{year}{2016}).
\newblock
\urldef\tempurl%
\url{https://www.propublica.org/article/machine-bias-risk-assessments-in-criminal-sentencing}
\showURL{%
\tempurl}


\bibitem[Asplund et~al\mbox{.}(2020)]%
        {asplund2020auditing}
\bibfield{author}{\bibinfo{person}{Joshua Asplund}, \bibinfo{person}{Motahhare
  Eslami}, \bibinfo{person}{Hari Sundaram}, \bibinfo{person}{Christian
  Sandvig}, {and} \bibinfo{person}{Karrie Karahalios}.}
  \bibinfo{year}{2020}\natexlab{}.
\newblock \showarticletitle{Auditing race and gender discrimination in online
  housing markets}. In \bibinfo{booktitle}{\emph{Proceedings of the
  International AAAI Conference on Web and Social Media}},
  Vol.~\bibinfo{volume}{14}. \bibinfo{pages}{24--35}.
\newblock


\bibitem[Attenberg et~al\mbox{.}(2015)]%
        {attenberg2015beat}
\bibfield{author}{\bibinfo{person}{Joshua Attenberg}, \bibinfo{person}{Panos
  Ipeirotis}, {and} \bibinfo{person}{Foster Provost}.}
  \bibinfo{year}{2015}\natexlab{}.
\newblock \showarticletitle{Beat the machine: Challenging humans to find a
  predictive model's “unknown unknowns”}.
\newblock \bibinfo{journal}{\emph{Journal of Data and Information Quality
  (JDIQ)}} \bibinfo{volume}{6}, \bibinfo{number}{1} (\bibinfo{year}{2015}),
  \bibinfo{pages}{1--17}.
\newblock


\bibitem[Attenberg et~al\mbox{.}(2011)]%
        {attenberg2011beat}
\bibfield{author}{\bibinfo{person}{Josh~M Attenberg},
  \bibinfo{person}{Pagagiotis~G Ipeirotis}, {and} \bibinfo{person}{Foster
  Provost}.} \bibinfo{year}{2011}\natexlab{}.
\newblock \showarticletitle{Beat the machine: Challenging workers to find the
  unknown unknowns}. In \bibinfo{booktitle}{\emph{Workshops at the Twenty-Fifth
  AAAI Conference on Artificial Intelligence}}.
\newblock


\bibitem[Barocas and Selbst(2016)]%
        {barocas2016big}
\bibfield{author}{\bibinfo{person}{Solon Barocas} {and}
  \bibinfo{person}{Andrew~D Selbst}.} \bibinfo{year}{2016}\natexlab{}.
\newblock \showarticletitle{Big data's disparate impact}.
\newblock \bibinfo{journal}{\emph{Calif. L. Rev.}}  \bibinfo{volume}{104}
  (\bibinfo{year}{2016}), \bibinfo{pages}{671}.
\newblock


\bibitem[Bellamy et~al\mbox{.}(2019)]%
        {bellamy2018ai}
\bibfield{author}{\bibinfo{person}{Rachel~KE Bellamy}, \bibinfo{person}{Kuntal
  Dey}, \bibinfo{person}{Michael Hind}, \bibinfo{person}{Samuel~C Hoffman},
  \bibinfo{person}{Stephanie Houde}, \bibinfo{person}{Kalapriya Kannan},
  \bibinfo{person}{Pranay Lohia}, \bibinfo{person}{Jacquelyn Martino},
  \bibinfo{person}{Sameep Mehta}, \bibinfo{person}{Aleksandra Mojsilovi{\'c}},
  {et~al\mbox{.}}} \bibinfo{year}{2019}\natexlab{}.
\newblock \showarticletitle{AI Fairness 360: An extensible toolkit for
  detecting and mitigating algorithmic bias}.
\newblock \bibinfo{journal}{\emph{IBM Journal of Research and Development}}
  \bibinfo{volume}{63}, \bibinfo{number}{4/5} (\bibinfo{year}{2019}),
  \bibinfo{pages}{4--1}.
\newblock


\bibitem[Bigham et~al\mbox{.}(2015)]%
        {bigham2015human}
\bibfield{author}{\bibinfo{person}{Jeffrey~P Bigham},
  \bibinfo{person}{Michael~S Bernstein}, {and} \bibinfo{person}{Eytan Adar}.}
  \bibinfo{year}{2015}\natexlab{}.
\newblock \showarticletitle{Human-computer interaction and collective
  intelligence}.
\newblock \bibinfo{journal}{\emph{Handbook of collective intelligence}}
  \bibinfo{volume}{57} (\bibinfo{year}{2015}).
\newblock


\bibitem[Bird(2020)]%
        {FairlearnAPI}
\bibfield{author}{\bibinfo{person}{Sarah Bird}.}
  \bibinfo{year}{2020}\natexlab{}.
\newblock \bibinfo{title}{Fairlearn API}.
\newblock
\newblock
\urldef\tempurl%
\url{https://fairlearn.github.io/v0.5.0/api_reference/fairlearn.datasets.html}
\showURL{%
\tempurl}


\bibitem[Bird et~al\mbox{.}(2020)]%
        {bird2020fairlearn}
\bibfield{author}{\bibinfo{person}{Sarah Bird}, \bibinfo{person}{Miro Dudík},
  \bibinfo{person}{Richard Edgar}, \bibinfo{person}{Brandon Horn},
  \bibinfo{person}{Roman Lutz}, \bibinfo{person}{Vanessa Milan},
  \bibinfo{person}{Mehrnoosh Sameki}, \bibinfo{person}{Hanna Wallach}, {and}
  \bibinfo{person}{Kathleen Walker}.} \bibinfo{year}{2020}\natexlab{}.
\newblock \bibinfo{booktitle}{\emph{Fairlearn: A toolkit for assessing and
  improving fairness in AI}}.
\newblock \bibinfo{type}{{T}echnical {R}eport} MSR-TR-2020-32.
  \bibinfo{institution}{Microsoft}.
\newblock
\urldef\tempurl%
\url{https://www.microsoft.com/en-us/research/publication/fairlearn-a-toolkit-for-assessing-and-improving-fairness-in-ai/}
\showURL{%
\tempurl}


\bibitem[Braun and Clarke(2019)]%
        {braun2019reflecting}
\bibfield{author}{\bibinfo{person}{Virginia Braun} {and}
  \bibinfo{person}{Victoria Clarke}.} \bibinfo{year}{2019}\natexlab{}.
\newblock \showarticletitle{Reflecting on reflexive thematic analysis}.
\newblock \bibinfo{journal}{\emph{Qualitative research in sport, exercise and
  health}} \bibinfo{volume}{11}, \bibinfo{number}{4} (\bibinfo{year}{2019}),
  \bibinfo{pages}{589--597}.
\newblock


\bibitem[Buolamwini and Gebru(2018)]%
        {buolamwini2018gender}
\bibfield{author}{\bibinfo{person}{Joy Buolamwini} {and}
  \bibinfo{person}{Timnit Gebru}.} \bibinfo{year}{2018}\natexlab{}.
\newblock \showarticletitle{Gender shades: Intersectional accuracy disparities
  in commercial gender classification}. In \bibinfo{booktitle}{\emph{Conference
  on fairness, accountability and transparency}}. PMLR,
  \bibinfo{pages}{77--91}.
\newblock


\bibitem[Cabrera et~al\mbox{.}(2021)]%
        {cabrera2021discovering}
\bibfield{author}{\bibinfo{person}{{\'A}ngel~Alexander Cabrera},
  \bibinfo{person}{Abraham~J Druck}, \bibinfo{person}{Jason~I Hong}, {and}
  \bibinfo{person}{Adam Perer}.} \bibinfo{year}{2021}\natexlab{}.
\newblock \showarticletitle{Discovering and validating ai errors with
  crowdsourced failure reports}.
\newblock \bibinfo{journal}{\emph{Proceedings of the ACM on Human-Computer
  Interaction}} \bibinfo{volume}{5}, \bibinfo{number}{CSCW2}
  (\bibinfo{year}{2021}), \bibinfo{pages}{1--22}.
\newblock


\bibitem[Callison-Burch and Dredze(2010)]%
        {callison2010creating}
\bibfield{author}{\bibinfo{person}{Chris Callison-Burch} {and}
  \bibinfo{person}{Mark Dredze}.} \bibinfo{year}{2010}\natexlab{}.
\newblock \showarticletitle{Creating speech and language data with amazon’s
  mechanical turk}. In \bibinfo{booktitle}{\emph{Proceedings of the NAACL HLT
  2010 workshop on creating speech and language data with Amazon’s Mechanical
  Turk}}. \bibinfo{pages}{1--12}.
\newblock


\bibitem[Campbell et~al\mbox{.}(2020)]%
        {campbell2020purposive}
\bibfield{author}{\bibinfo{person}{Steve Campbell}, \bibinfo{person}{Melanie
  Greenwood}, \bibinfo{person}{Sarah Prior}, \bibinfo{person}{Toniele Shearer},
  \bibinfo{person}{Kerrie Walkem}, \bibinfo{person}{Sarah Young},
  \bibinfo{person}{Danielle Bywaters}, {and} \bibinfo{person}{Kim Walker}.}
  \bibinfo{year}{2020}\natexlab{}.
\newblock \showarticletitle{Purposive sampling: complex or simple? Research
  case examples}.
\newblock \bibinfo{journal}{\emph{Journal of research in Nursing}}
  \bibinfo{volume}{25}, \bibinfo{number}{8} (\bibinfo{year}{2020}),
  \bibinfo{pages}{652--661}.
\newblock


\bibitem[Chen et~al\mbox{.}(2019)]%
        {chen2019fairness}
\bibfield{author}{\bibinfo{person}{Jiahao Chen}, \bibinfo{person}{Nathan
  Kallus}, \bibinfo{person}{Xiaojie Mao}, \bibinfo{person}{Geoffry Svacha},
  {and} \bibinfo{person}{Madeleine Udell}.} \bibinfo{year}{2019}\natexlab{}.
\newblock \showarticletitle{Fairness under unawareness: Assessing disparity
  when protected class is unobserved}. In \bibinfo{booktitle}{\emph{Proceedings
  of the conference on fairness, accountability, and transparency}}.
  \bibinfo{pages}{339--348}.
\newblock


\bibitem[Chowdhury and Williams(2021)]%
        {chowdhury2021introducing}
\bibfield{author}{\bibinfo{person}{Rumman Chowdhury} {and}
  \bibinfo{person}{Jutta Williams}.} \bibinfo{year}{2021}\natexlab{}.
\newblock \showarticletitle{Introducing Twitter’s first algorithmic bias
  bounty challenge}.
\newblock \bibinfo{journal}{\emph{URl: https://blog. twitter.
  com/engineering/en\_us/topics/insights/2021/algorithmic-bias-bountychallenge}}
  (\bibinfo{year}{2021}).
\newblock


\bibitem[Cramer et~al\mbox{.}(2018)]%
        {cramer2018assessing}
\bibfield{author}{\bibinfo{person}{Henriette Cramer}, \bibinfo{person}{Jean
  Garcia-Gathright}, \bibinfo{person}{Aaron Springer}, {and}
  \bibinfo{person}{Sravana Reddy}.} \bibinfo{year}{2018}\natexlab{}.
\newblock \showarticletitle{Assessing and addressing algorithmic bias in
  practice}.
\newblock \bibinfo{journal}{\emph{Interactions}} \bibinfo{volume}{25},
  \bibinfo{number}{6} (\bibinfo{year}{2018}), \bibinfo{pages}{58--63}.
\newblock


\bibitem[Crawford(2017)]%
        {crawford2017trouble}
\bibfield{author}{\bibinfo{person}{Kate Crawford}.}
  \bibinfo{year}{2017}\natexlab{}.
\newblock \showarticletitle{The trouble with bias}. In
  \bibinfo{booktitle}{\emph{Conference on Neural Information Processing
  Systems, invited speaker}}.
\newblock


\bibitem[Deng et~al\mbox{.}(2022)]%
        {deng2022exploring}
\bibfield{author}{\bibinfo{person}{Wesley~Hanwen Deng}, \bibinfo{person}{Manish
  Nagireddy}, \bibinfo{person}{Michelle Seng~Ah Lee}, \bibinfo{person}{Jatinder
  Singh}, \bibinfo{person}{Zhiwei~Steven Wu}, \bibinfo{person}{Kenneth
  Holstein}, {and} \bibinfo{person}{Haiyi Zhu}.}
  \bibinfo{year}{2022}\natexlab{}.
\newblock \showarticletitle{Exploring {How} {Machine} {Learning}
  {Practitioners} ({Try} {To}) {Use} {Fairness} {Toolkits}}. In
  \bibinfo{booktitle}{\emph{2022 {ACM} {Conference} on {Fairness},
  {Accountability}, and {Transparency}}}. \bibinfo{publisher}{ACM},
  \bibinfo{address}{Seoul Republic of Korea}, \bibinfo{pages}{473--484}.
\newblock
\showISBNx{978-1-4503-9352-2}
\urldef\tempurl%
\url{https://doi.org/10.1145/3531146.3533113}
\showDOI{\tempurl}


\bibitem[DeVos et~al\mbox{.}(2022)]%
        {DeVos2022TowardUA}
\bibfield{author}{\bibinfo{person}{Alicia DeVos}, \bibinfo{person}{Aditi
  Dhabalia}, \bibinfo{person}{Hong Shen}, \bibinfo{person}{Kenneth Holstein},
  {and} \bibinfo{person}{Motahhare Eslami}.} \bibinfo{year}{2022}\natexlab{}.
\newblock \showarticletitle{Toward User-Driven Algorithm Auditing:
  Investigating users’ strategies for uncovering harmful algorithmic
  behavior}.
\newblock \bibinfo{journal}{\emph{CHI Conference on Human Factors in Computing
  Systems}} (\bibinfo{year}{2022}).
\newblock


\bibitem[Dosono and Semaan(2019)]%
        {dosono2019moderation}
\bibfield{author}{\bibinfo{person}{Bryan Dosono} {and} \bibinfo{person}{Bryan
  Semaan}.} \bibinfo{year}{2019}\natexlab{}.
\newblock \showarticletitle{Moderation practices as emotional labor in
  sustaining online communities: The case of AAPI identity work on Reddit}. In
  \bibinfo{booktitle}{\emph{Proceedings of the 2019 CHI conference on human
  factors in computing systems}}. \bibinfo{pages}{1--13}.
\newblock


\bibitem[Eslami et~al\mbox{.}(2016)]%
        {eslami2016first}
\bibfield{author}{\bibinfo{person}{Motahhare Eslami}, \bibinfo{person}{Karrie
  Karahalios}, \bibinfo{person}{Christian Sandvig}, \bibinfo{person}{Kristen
  Vaccaro}, \bibinfo{person}{Aimee Rickman}, \bibinfo{person}{Kevin Hamilton},
  {and} \bibinfo{person}{Alex Kirlik}.} \bibinfo{year}{2016}\natexlab{}.
\newblock \bibinfo{booktitle}{\emph{First I "like" It, Then I Hide It: Folk
  Theories of Social Feeds}}.
\newblock \bibinfo{publisher}{Association for Computing Machinery},
  \bibinfo{address}{New York, NY, USA}, \bibinfo{pages}{2371–2382}.
\newblock
\showISBNx{9781450333627}
\urldef\tempurl%
\url{https://doi.org/10.1145/2858036.2858494}
\showURL{%
\tempurl}


\bibitem[Eslami et~al\mbox{.}(2015)]%
        {eslami2015i}
\bibfield{author}{\bibinfo{person}{Motahhare Eslami}, \bibinfo{person}{Aimee
  Rickman}, \bibinfo{person}{Kristen Vaccaro}, \bibinfo{person}{Amirhossein
  Aleyasen}, \bibinfo{person}{Andy Vuong}, \bibinfo{person}{Karrie Karahalios},
  \bibinfo{person}{Kevin Hamilton}, {and} \bibinfo{person}{Christian Sandvig}.}
  \bibinfo{year}{2015}\natexlab{}.
\newblock \showarticletitle{"I Always Assumed That I Wasn't Really That Close
  to [Her]": Reasoning about Invisible Algorithms in News Feeds}. In
  \bibinfo{booktitle}{\emph{Proceedings of the 33rd Annual ACM Conference on
  Human Factors in Computing Systems}} (Seoul, Republic of Korea)
  \emph{(\bibinfo{series}{CHI '15})}. \bibinfo{publisher}{Association for
  Computing Machinery}, \bibinfo{address}{New York, NY, USA},
  \bibinfo{pages}{153–162}.
\newblock
\showISBNx{9781450331456}
\urldef\tempurl%
\url{https://doi.org/10.1145/2702123.2702556}
\showDOI{\tempurl}


\bibitem[Eslami et~al\mbox{.}(2017a)]%
        {eslami2017biased}
\bibfield{author}{\bibinfo{person}{Motahhare Eslami}, \bibinfo{person}{Kristen
  Vaccaro}, \bibinfo{person}{Karrie Karahalios}, {and} \bibinfo{person}{Kevin
  Hamilton}.} \bibinfo{year}{2017}\natexlab{a}.
\newblock \showarticletitle{Be careful; Things can be worse than they appear -
  Understanding biased algorithms and users' behavior around them in rating
  platforms"}.
\newblock  (\bibinfo{year}{2017}), \bibinfo{pages}{62--71}.
\newblock
\newblock
\shownote{Funding Information: This work was funded by NSF grant CHS-1564041.
  Publisher Copyright: {\textcopyright} Copyright 2017, Association for the
  Advancement of Artificial Intelligence (www.aaai.org). All rights reserved.;
  11th International Conference on Web and Social Media, ICWSM 2017 ;
  Conference date: 15-05-2017 Through 18-05-2017}.


\bibitem[Eslami et~al\mbox{.}(2017b)]%
        {eslami2017careful}
\bibfield{author}{\bibinfo{person}{Motahhare Eslami}, \bibinfo{person}{Kristen
  Vaccaro}, \bibinfo{person}{Karrie Karahalios}, {and} \bibinfo{person}{Kevin
  Hamilton}.} \bibinfo{year}{2017}\natexlab{b}.
\newblock \showarticletitle{“Be careful; things can be worse than they
  appear”: Understanding Biased Algorithms and Users’ Behavior around Them
  in Rating Platforms}. In \bibinfo{booktitle}{\emph{Proceedings of the
  International AAAI Conference on Web and Social Media}},
  Vol.~\bibinfo{volume}{11}.
\newblock


\bibitem[Eslami et~al\mbox{.}(2019)]%
        {eslami2019opacity}
\bibfield{author}{\bibinfo{person}{Motahhare Eslami}, \bibinfo{person}{Kristen
  Vaccaro}, \bibinfo{person}{Min~Kyung Lee}, \bibinfo{person}{Amit Elazari
  Bar~On}, \bibinfo{person}{Eric Gilbert}, {and} \bibinfo{person}{Karrie
  Karahalios}.} \bibinfo{year}{2019}\natexlab{}.
\newblock \showarticletitle{User Attitudes towards Algorithmic Opacity and
  Transparency in Online Reviewing Platforms}.
\newblock  (\bibinfo{year}{2019}), \bibinfo{pages}{1–14}.
\newblock
\showISBNx{9781450359702}
\urldef\tempurl%
\url{https://doi.org/10.1145/3290605.3300724}
\showURL{%
\tempurl}


\bibitem[Eubanks(2018)]%
        {eubanks2018automating}
\bibfield{author}{\bibinfo{person}{Virginia Eubanks}.}
  \bibinfo{year}{2018}\natexlab{}.
\newblock \bibinfo{booktitle}{\emph{Automating inequality: How high-tech tools
  profile, police, and punish the poor}}.
\newblock \bibinfo{publisher}{St. Martin's Press}.
\newblock


\bibitem[Evenson(2006)]%
        {evenson2006directed}
\bibfield{author}{\bibinfo{person}{Shelley Evenson}.}
  \bibinfo{year}{2006}\natexlab{}.
\newblock \showarticletitle{Directed storytelling: Interpreting experience for
  design}.
\newblock \bibinfo{journal}{\emph{Design Studies: Theory and research in
  graphic design}} (\bibinfo{year}{2006}), \bibinfo{pages}{231--240}.
\newblock


\bibitem[Friedman and Nissenbaum(1996)]%
        {friedman1996bias}
\bibfield{author}{\bibinfo{person}{Batya Friedman} {and} \bibinfo{person}{Helen
  Nissenbaum}.} \bibinfo{year}{1996}\natexlab{}.
\newblock \showarticletitle{Bias in Computer Systems}.
\newblock \bibinfo{journal}{\emph{ACM Trans. Inf. Syst.}} \bibinfo{volume}{14},
  \bibinfo{number}{3} (\bibinfo{date}{July} \bibinfo{year}{1996}),
  \bibinfo{pages}{330–347}.
\newblock
\showISSN{1046-8188}
\urldef\tempurl%
\url{https://doi.org/10.1145/230538.230561}
\showDOI{\tempurl}


\bibitem[Gordon et~al\mbox{.}(2022)]%
        {gordon2022jury}
\bibfield{author}{\bibinfo{person}{Mitchell~L Gordon},
  \bibinfo{person}{Michelle~S Lam}, \bibinfo{person}{Joon~Sung Park},
  \bibinfo{person}{Kayur Patel}, \bibinfo{person}{Jeff Hancock},
  \bibinfo{person}{Tatsunori Hashimoto}, {and} \bibinfo{person}{Michael~S
  Bernstein}.} \bibinfo{year}{2022}\natexlab{}.
\newblock \showarticletitle{Jury learning: Integrating dissenting voices into
  machine learning models}. In \bibinfo{booktitle}{\emph{CHI Conference on
  Human Factors in Computing Systems}}. \bibinfo{pages}{1--19}.
\newblock


\bibitem[Groce et~al\mbox{.}(2013)]%
        {groce2013you}
\bibfield{author}{\bibinfo{person}{Alex Groce}, \bibinfo{person}{Todd Kulesza},
  \bibinfo{person}{Chaoqiang Zhang}, \bibinfo{person}{Shalini Shamasunder},
  \bibinfo{person}{Margaret Burnett}, \bibinfo{person}{Weng-Keen Wong},
  \bibinfo{person}{Simone Stumpf}, \bibinfo{person}{Shubhomoy Das},
  \bibinfo{person}{Amber Shinsel}, \bibinfo{person}{Forrest Bice},
  {et~al\mbox{.}}} \bibinfo{year}{2013}\natexlab{}.
\newblock \showarticletitle{You are the only possible oracle: Effective test
  selection for end users of interactive machine learning systems}.
\newblock \bibinfo{journal}{\emph{IEEE Transactions on Software Engineering}}
  \bibinfo{volume}{40}, \bibinfo{number}{3} (\bibinfo{year}{2013}),
  \bibinfo{pages}{307--323}.
\newblock


\bibitem[Hannak et~al\mbox{.}(2014)]%
        {hannak2014measuring}
\bibfield{author}{\bibinfo{person}{Aniko Hannak}, \bibinfo{person}{Gary
  Soeller}, \bibinfo{person}{David Lazer}, \bibinfo{person}{Alan Mislove},
  {and} \bibinfo{person}{Christo Wilson}.} \bibinfo{year}{2014}\natexlab{}.
\newblock \showarticletitle{Measuring price discrimination and steering on
  e-commerce web sites}. In \bibinfo{booktitle}{\emph{Proceedings of the 2014
  Conference on Internet Measurement Conference}}. \bibinfo{pages}{305--318}.
\newblock


\bibitem[Harrington et~al\mbox{.}(2019)]%
        {harrington2019deconstructing}
\bibfield{author}{\bibinfo{person}{Christina Harrington},
  \bibinfo{person}{Sheena Erete}, {and} \bibinfo{person}{Anne~Marie Piper}.}
  \bibinfo{year}{2019}\natexlab{}.
\newblock \showarticletitle{Deconstructing community-based collaborative
  design: Towards more equitable participatory design engagements}.
\newblock \bibinfo{journal}{\emph{Proceedings of the ACM on Human-Computer
  Interaction}} \bibinfo{volume}{3}, \bibinfo{number}{CSCW}
  (\bibinfo{year}{2019}), \bibinfo{pages}{1--25}.
\newblock


\bibitem[Harris(2022)]%
        {Facebook_Ban}
\bibfield{author}{\bibinfo{person}{Jamie Harris}.}
  \bibinfo{year}{2022}\natexlab{}.
\newblock \bibinfo{title}{Facebook forced to ban its AI after it ‘revealed
  how to make napalm and made racist comments}.
\newblock
\newblock
\urldef\tempurl%
\url{https://www.the-sun.com/tech/6729391/meta-withdraws-ai-galactica-controversy/}
\showURL{%
\tempurl}


\bibitem[Holstein et~al\mbox{.}(2020)]%
        {holstein2020replay}
\bibfield{author}{\bibinfo{person}{Kenneth Holstein}, \bibinfo{person}{Erik
  Harpstead}, \bibinfo{person}{Rebecca Gulotta}, {and} \bibinfo{person}{Jodi
  Forlizzi}.} \bibinfo{year}{2020}\natexlab{}.
\newblock \showarticletitle{Replay enactments: Exploring possible futures
  through historical data}. In \bibinfo{booktitle}{\emph{Proceedings of the
  2020 ACM Designing Interactive Systems Conference}}.
  \bibinfo{pages}{1607--1618}.
\newblock


\bibitem[Holstein et~al\mbox{.}(2019a)]%
        {holstein2019co}
\bibfield{author}{\bibinfo{person}{Kenneth Holstein}, \bibinfo{person}{Bruce~M
  McLaren}, {and} \bibinfo{person}{Vincent Aleven}.}
  \bibinfo{year}{2019}\natexlab{a}.
\newblock \showarticletitle{Designing for complementarity: Teacher and student
  needs for orchestration support in AI-enhanced classrooms}.
\newblock  (\bibinfo{year}{2019}), \bibinfo{pages}{157--171}.
\newblock


\bibitem[Holstein et~al\mbox{.}(2019b)]%
        {holstein2019improving}
\bibfield{author}{\bibinfo{person}{Kenneth Holstein}, \bibinfo{person}{Jennifer
  Wortman~Vaughan}, \bibinfo{person}{Hal Daum{\'e}~III}, \bibinfo{person}{Miro
  Dudik}, {and} \bibinfo{person}{Hanna Wallach}.}
  \bibinfo{year}{2019}\natexlab{b}.
\newblock \showarticletitle{Improving fairness in machine learning systems:
  What do industry practitioners need?}. In
  \bibinfo{booktitle}{\emph{Proceedings of the 2019 CHI Conference on Human
  Factors in Computing Systems}}. \bibinfo{pages}{1--16}.
\newblock


\bibitem[Hossain(2012)]%
        {hossain2012users}
\bibfield{author}{\bibinfo{person}{Mokter Hossain}.}
  \bibinfo{year}{2012}\natexlab{}.
\newblock \showarticletitle{Users' motivation to participate in online
  crowdsourcing platforms}. In \bibinfo{booktitle}{\emph{2012 International
  Conference on Innovation Management and Technology Research}}. IEEE,
  \bibinfo{pages}{310--315}.
\newblock


\bibitem[Hsu et~al\mbox{.}(2022)]%
        {hsu2022empowering}
\bibfield{author}{\bibinfo{person}{Yen-Chia Hsu}, \bibinfo{person}{Himanshu
  Verma}, \bibinfo{person}{Andrea Mauri}, \bibinfo{person}{Illah Nourbakhsh},
  \bibinfo{person}{Alessandro Bozzon}, {et~al\mbox{.}}}
  \bibinfo{year}{2022}\natexlab{}.
\newblock \showarticletitle{Empowering local communities using artificial
  intelligence}.
\newblock \bibinfo{journal}{\emph{Patterns}} \bibinfo{volume}{3},
  \bibinfo{number}{3} (\bibinfo{year}{2022}), \bibinfo{pages}{100449}.
\newblock


\bibitem[Irani and Silberman(2013)]%
        {irani2013turkopticon}
\bibfield{author}{\bibinfo{person}{Lilly~C Irani} {and} \bibinfo{person}{M~Six
  Silberman}.} \bibinfo{year}{2013}\natexlab{}.
\newblock \showarticletitle{Turkopticon: Interrupting worker invisibility in
  amazon mechanical turk}. In \bibinfo{booktitle}{\emph{Proceedings of the
  SIGCHI conference on human factors in computing systems}}.
  \bibinfo{pages}{611--620}.
\newblock


\bibitem[Kaufmann et~al\mbox{.}(2011)]%
        {Buolamwini2019hearing}
\bibfield{author}{\bibinfo{person}{Nicolas Kaufmann}, \bibinfo{person}{Thimo
  Schulze}, {and} \bibinfo{person}{Daniel Veit}.}
  \bibinfo{year}{2011}\natexlab{}.
\newblock \showarticletitle{More than fun and money. worker motivation in
  crowdsourcing--a study on mechanical turk}.
\newblock  (\bibinfo{year}{2011}).
\newblock


\bibitem[Kaur et~al\mbox{.}(2020)]%
        {kaur2020interpreting}
\bibfield{author}{\bibinfo{person}{Harmanpreet Kaur}, \bibinfo{person}{Harsha
  Nori}, \bibinfo{person}{Samuel Jenkins}, \bibinfo{person}{Rich Caruana},
  \bibinfo{person}{Hanna Wallach}, {and} \bibinfo{person}{Jennifer
  Wortman~Vaughan}.} \bibinfo{year}{2020}\natexlab{}.
\newblock \showarticletitle{Interpreting interpretability: understanding data
  scientists' use of interpretability tools for machine learning}. In
  \bibinfo{booktitle}{\emph{Proceedings of the 2020 CHI conference on human
  factors in computing systems}}. \bibinfo{pages}{1--14}.
\newblock


\bibitem[Kiela et~al\mbox{.}(2021)]%
        {kiela2021dynabench}
\bibfield{author}{\bibinfo{person}{Douwe Kiela}, \bibinfo{person}{Max Bartolo},
  \bibinfo{person}{Yixin Nie}, \bibinfo{person}{Divyansh Kaushik},
  \bibinfo{person}{Atticus Geiger}, \bibinfo{person}{Zhengxuan Wu},
  \bibinfo{person}{Bertie Vidgen}, \bibinfo{person}{Grusha Prasad},
  \bibinfo{person}{Amanpreet Singh}, \bibinfo{person}{Pratik Ringshia},
  {et~al\mbox{.}}} \bibinfo{year}{2021}\natexlab{}.
\newblock \showarticletitle{Dynabench: Rethinking benchmarking in NLP}.
\newblock \bibinfo{journal}{\emph{arXiv preprint arXiv:2104.14337}}
  (\bibinfo{year}{2021}).
\newblock


\bibitem[Kittur et~al\mbox{.}(2013a)]%
        {kittur2013future}
\bibfield{author}{\bibinfo{person}{Aniket Kittur}, \bibinfo{person}{Jeffrey~V
  Nickerson}, \bibinfo{person}{Michael Bernstein}, \bibinfo{person}{Elizabeth
  Gerber}, \bibinfo{person}{Aaron Shaw}, \bibinfo{person}{John Zimmerman},
  \bibinfo{person}{Matt Lease}, {and} \bibinfo{person}{John Horton}.}
  \bibinfo{year}{2013}\natexlab{a}.
\newblock \showarticletitle{The future of crowd work}. In
  \bibinfo{booktitle}{\emph{Proceedings of the 2013 conference on Computer
  supported cooperative work}}. \bibinfo{pages}{1301--1318}.
\newblock


\bibitem[Kittur et~al\mbox{.}(2013b)]%
        {Kittur2013TheFO}
\bibfield{author}{\bibinfo{person}{Aniket Kittur}, \bibinfo{person}{Jeffrey~V.
  Nickerson}, \bibinfo{person}{Michael~S. Bernstein},
  \bibinfo{person}{Elizabeth Gerber}, \bibinfo{person}{Aaron~D. Shaw},
  \bibinfo{person}{John Zimmerman}, \bibinfo{person}{Matthew Lease}, {and}
  \bibinfo{person}{John~Joseph Horton}.} \bibinfo{year}{2013}\natexlab{b}.
\newblock \showarticletitle{The future of crowd work}.
\newblock \bibinfo{journal}{\emph{Proceedings of the 2013 conference on
  Computer supported cooperative work}} (\bibinfo{year}{2013}).
\newblock


\bibitem[Kraut and Resnick(2011)]%
        {kraut2011encouraging}
\bibfield{author}{\bibinfo{person}{Robert~E Kraut} {and} \bibinfo{person}{Paul
  Resnick}.} \bibinfo{year}{2011}\natexlab{}.
\newblock \showarticletitle{Encouraging contribution to online communities}.
\newblock \bibinfo{journal}{\emph{Building successful online communities:
  Evidence-based social design}} (\bibinfo{year}{2011}),
  \bibinfo{pages}{21--76}.
\newblock


\bibitem[Kroll et~al\mbox{.}(2016)]%
        {kroll2016accountable}
\bibfield{author}{\bibinfo{person}{Joshua~A Kroll}, \bibinfo{person}{Solon
  Barocas}, \bibinfo{person}{Edward~W Felten}, \bibinfo{person}{Joel~R
  Reidenberg}, \bibinfo{person}{David~G Robinson}, {and}
  \bibinfo{person}{Harlan Yu}.} \bibinfo{year}{2016}\natexlab{}.
\newblock \showarticletitle{Accountable algorithms}.
\newblock \bibinfo{journal}{\emph{U. Pa. L. Rev.}}  \bibinfo{volume}{165}
  (\bibinfo{year}{2016}), \bibinfo{pages}{633}.
\newblock


\bibitem[Lam et~al\mbox{.}(2022)]%
        {lam2022enduser}
\bibfield{author}{\bibinfo{person}{Michelle~S. Lam},
  \bibinfo{person}{Mitchell~L. Gordon}, \bibinfo{person}{Dana\"{e} Metaxa},
  \bibinfo{person}{Jeffrey~T. Hancock}, \bibinfo{person}{James~A. Landay},
  {and} \bibinfo{person}{Michael~S. Bernstein}.}
  \bibinfo{year}{2022}\natexlab{}.
\newblock \showarticletitle{End-User Audits: A System Empowering Communities to
  Lead Large-Scale Investigations of Harmful Algorithmic Behavior}.
\newblock \bibinfo{journal}{\emph{Proc. ACM Hum.-Comput. Interact.}}
  \bibinfo{volume}{6}, \bibinfo{number}{CSCW2}, Article
  \bibinfo{articleno}{512} (\bibinfo{date}{Nov} \bibinfo{year}{2022}),
  \bibinfo{numpages}{34}~pages.
\newblock
\urldef\tempurl%
\url{https://doi.org/10.1145/3555625}
\showDOI{\tempurl}


\bibitem[Lee and Singh(2020)]%
        {lee2020landscape}
\bibfield{author}{\bibinfo{person}{Michelle Seng~Ah Lee} {and}
  \bibinfo{person}{Jatinder Singh}.} \bibinfo{year}{2020}\natexlab{}.
\newblock \showarticletitle{The Landscape and Gaps in Open Source Fairness
  Toolkits}.
\newblock \bibinfo{journal}{\emph{Available at SSRN}} (\bibinfo{year}{2020}).
\newblock


\bibitem[Ling et~al\mbox{.}(2005)]%
        {ling2005using}
\bibfield{author}{\bibinfo{person}{Kimberly Ling}, \bibinfo{person}{Gerard
  Beenen}, \bibinfo{person}{Pamela Ludford}, \bibinfo{person}{Xiaoqing Wang},
  \bibinfo{person}{Klarissa Chang}, \bibinfo{person}{Xin Li},
  \bibinfo{person}{Dan Cosley}, \bibinfo{person}{Dan Frankowski},
  \bibinfo{person}{Loren Terveen}, \bibinfo{person}{Al~Mamunur Rashid},
  {et~al\mbox{.}}} \bibinfo{year}{2005}\natexlab{}.
\newblock \showarticletitle{Using social psychology to motivate contributions
  to online communities}.
\newblock \bibinfo{journal}{\emph{Journal of Computer-Mediated Communication}}
  \bibinfo{volume}{10}, \bibinfo{number}{4} (\bibinfo{year}{2005}),
  \bibinfo{pages}{00--00}.
\newblock


\bibitem[Madaio et~al\mbox{.}(2021)]%
        {madaio2021assessing}
\bibfield{author}{\bibinfo{person}{Michael Madaio}, \bibinfo{person}{Lisa
  Egede}, \bibinfo{person}{Hariharan Subramonyam},
  \bibinfo{person}{Jennifer~Wortman Vaughan}, {and} \bibinfo{person}{Hanna
  Wallach}.} \bibinfo{year}{2021}\natexlab{}.
\newblock \showarticletitle{Assessing the Fairness of AI Systems: AI
  Practitioners' Processes, Challenges, and Needs for Support}.
\newblock \bibinfo{journal}{\emph{arXiv preprint arXiv:2112.05675}}
  (\bibinfo{year}{2021}).
\newblock


\bibitem[Madaio et~al\mbox{.}(2020)]%
        {madaio2020co}
\bibfield{author}{\bibinfo{person}{Michael~A Madaio}, \bibinfo{person}{Luke
  Stark}, \bibinfo{person}{Jennifer Wortman~Vaughan}, {and}
  \bibinfo{person}{Hanna Wallach}.} \bibinfo{year}{2020}\natexlab{}.
\newblock \showarticletitle{Co-designing checklists to understand
  organizational challenges and opportunities around fairness in ai}. In
  \bibinfo{booktitle}{\emph{Proceedings of the 2020 CHI Conference on Human
  Factors in Computing Systems}}. \bibinfo{pages}{1--14}.
\newblock


\bibitem[Malladi and Subramanian(2019)]%
        {malladi2019bug}
\bibfield{author}{\bibinfo{person}{Suresh~S Malladi} {and}
  \bibinfo{person}{Hemang~C Subramanian}.} \bibinfo{year}{2019}\natexlab{}.
\newblock \showarticletitle{Bug bounty programs for cybersecurity: Practices,
  issues, and recommendations}.
\newblock \bibinfo{journal}{\emph{IEEE Software}} \bibinfo{volume}{37},
  \bibinfo{number}{1} (\bibinfo{year}{2019}), \bibinfo{pages}{31--39}.
\newblock


\bibitem[McDonald et~al\mbox{.}(2019)]%
        {mcdonald2019reliability}
\bibfield{author}{\bibinfo{person}{Nora McDonald}, \bibinfo{person}{Sarita
  Schoenebeck}, {and} \bibinfo{person}{Andrea Forte}.}
  \bibinfo{year}{2019}\natexlab{}.
\newblock \showarticletitle{Reliability and inter-rater reliability in
  qualitative research: Norms and guidelines for CSCW and HCI practice}.
\newblock \bibinfo{journal}{\emph{Proceedings of the ACM on human-computer
  interaction}} \bibinfo{volume}{3}, \bibinfo{number}{CSCW}
  (\bibinfo{year}{2019}), \bibinfo{pages}{1--23}.
\newblock


\bibitem[Metaxa et~al\mbox{.}(2021)]%
        {metaxa2021auditing}
\bibfield{author}{\bibinfo{person}{Dana{\"e} Metaxa},
  \bibinfo{person}{Joon~Sung Park}, \bibinfo{person}{Ronald~E Robertson},
  \bibinfo{person}{Karrie Karahalios}, \bibinfo{person}{Christo Wilson},
  \bibinfo{person}{Jeff Hancock}, \bibinfo{person}{Christian Sandvig},
  {et~al\mbox{.}}} \bibinfo{year}{2021}\natexlab{}.
\newblock \showarticletitle{Auditing algorithms: Understanding algorithmic
  systems from the outside in}.
\newblock \bibinfo{journal}{\emph{Foundations and Trends{\textregistered} in
  Human--Computer Interaction}} \bibinfo{volume}{14}, \bibinfo{number}{4}
  (\bibinfo{year}{2021}), \bibinfo{pages}{272--344}.
\newblock


\bibitem[Mitchell et~al\mbox{.}(2019)]%
        {mitchell2019model}
\bibfield{author}{\bibinfo{person}{Margaret Mitchell}, \bibinfo{person}{Simone
  Wu}, \bibinfo{person}{Andrew Zaldivar}, \bibinfo{person}{Parker Barnes},
  \bibinfo{person}{Lucy Vasserman}, \bibinfo{person}{Ben Hutchinson},
  \bibinfo{person}{Elena Spitzer}, \bibinfo{person}{Inioluwa~Deborah Raji},
  {and} \bibinfo{person}{Timnit Gebru}.} \bibinfo{year}{2019}\natexlab{}.
\newblock \showarticletitle{Model cards for model reporting}. In
  \bibinfo{booktitle}{\emph{Proceedings of the conference on fairness,
  accountability, and transparency}}. \bibinfo{pages}{220--229}.
\newblock


\bibitem[Noble(2018)]%
        {noble2018algorithms}
\bibfield{author}{\bibinfo{person}{Safiya~Umoja Noble}.}
  \bibinfo{year}{2018}\natexlab{}.
\newblock \bibinfo{booktitle}{\emph{Algorithms of oppression: How search
  engines reinforce racism}}.
\newblock \bibinfo{publisher}{NYU Press}.
\newblock


\bibitem[Nushi et~al\mbox{.}(2018)]%
        {nushi2018towards}
\bibfield{author}{\bibinfo{person}{Besmira Nushi}, \bibinfo{person}{Ece Kamar},
  {and} \bibinfo{person}{Eric Horvitz}.} \bibinfo{year}{2018}\natexlab{}.
\newblock \showarticletitle{Towards accountable ai: Hybrid human-machine
  analyses for characterizing system failure}. In
  \bibinfo{booktitle}{\emph{Proceedings of the AAAI Conference on Human
  Computation and Crowdsourcing}}, Vol.~\bibinfo{volume}{6}.
  \bibinfo{pages}{126--135}.
\newblock


\bibitem[Ochigame and Ye(2021)]%
        {ochigame2021search}
\bibfield{author}{\bibinfo{person}{Rodrigo Ochigame} {and}
  \bibinfo{person}{Katherine Ye}.} \bibinfo{year}{2021}\natexlab{}.
\newblock \showarticletitle{Search Atlas: Visualizing Divergent Search Results
  Across Geopolitical Borders}. In \bibinfo{booktitle}{\emph{Designing
  Interactive Systems Conference 2021}}. \bibinfo{pages}{1970--1983}.
\newblock


\bibitem[Pistilli(2022)]%
        {HuggingFace}
\bibfield{author}{\bibinfo{person}{Giada Pistilli}.}
  \bibinfo{year}{2022}\natexlab{}.
\newblock \bibinfo{title}{HuggingFace announcedthe new feature to flag any
  Model, Dataset, or Space on the Hub}.
\newblock
\newblock
\urldef\tempurl%
\url{https://twitter.com/GiadaPistilli/status/1571865167092396033?s=20&t=LRhhEu63s6ftPmtZdfz8Cw}
\showURL{%
\tempurl}


\bibitem[Prates et~al\mbox{.}(2020)]%
        {prates2020assessing}
\bibfield{author}{\bibinfo{person}{Marcelo~OR Prates}, \bibinfo{person}{Pedro~H
  Avelar}, {and} \bibinfo{person}{Lu{\'\i}s~C Lamb}.}
  \bibinfo{year}{2020}\natexlab{}.
\newblock \showarticletitle{Assessing gender bias in machine translation: a
  case study with google translate}.
\newblock \bibinfo{journal}{\emph{Neural Computing and Applications}}
  \bibinfo{volume}{32}, \bibinfo{number}{10} (\bibinfo{year}{2020}),
  \bibinfo{pages}{6363--6381}.
\newblock


\bibitem[Raji and Buolamwini(2019)]%
        {raji2019actionable}
\bibfield{author}{\bibinfo{person}{Inioluwa~Deborah Raji} {and}
  \bibinfo{person}{Joy Buolamwini}.} \bibinfo{year}{2019}\natexlab{}.
\newblock \showarticletitle{Actionable auditing: Investigating the impact of
  publicly naming biased performance results of commercial AI products}. In
  \bibinfo{booktitle}{\emph{Proceedings of the 2019 AAAI/ACM Conference on AI,
  Ethics, and Society}}. \bibinfo{pages}{429--435}.
\newblock


\bibitem[Raji et~al\mbox{.}(2020)]%
        {raji2020closing}
\bibfield{author}{\bibinfo{person}{Inioluwa~Deborah Raji},
  \bibinfo{person}{Andrew Smart}, \bibinfo{person}{Rebecca~N White},
  \bibinfo{person}{Margaret Mitchell}, \bibinfo{person}{Timnit Gebru},
  \bibinfo{person}{Ben Hutchinson}, \bibinfo{person}{Jamila Smith-Loud},
  \bibinfo{person}{Daniel Theron}, {and} \bibinfo{person}{Parker Barnes}.}
  \bibinfo{year}{2020}\natexlab{}.
\newblock \showarticletitle{Closing the AI accountability gap: Defining an
  end-to-end framework for internal algorithmic auditing}. In
  \bibinfo{booktitle}{\emph{Proceedings of the 2020 conference on fairness,
  accountability, and transparency}}. \bibinfo{pages}{33--44}.
\newblock


\bibitem[Raji et~al\mbox{.}(2022)]%
        {raji2022outsider}
\bibfield{author}{\bibinfo{person}{Inioluwa~Deborah Raji},
  \bibinfo{person}{Peggy Xu}, \bibinfo{person}{Colleen Honigsberg}, {and}
  \bibinfo{person}{Daniel Ho}.} \bibinfo{year}{2022}\natexlab{}.
\newblock \showarticletitle{Outsider Oversight: Designing a Third Party Audit
  Ecosystem for AI Governance}. In \bibinfo{booktitle}{\emph{Proceedings of the
  2022 AAAI/ACM Conference on AI, Ethics, and Society}}.
  \bibinfo{pages}{557--571}.
\newblock


\bibitem[Rakova et~al\mbox{.}(2021)]%
        {rakova2021responsible}
\bibfield{author}{\bibinfo{person}{Bogdana Rakova}, \bibinfo{person}{Jingying
  Yang}, \bibinfo{person}{Henriette Cramer}, {and} \bibinfo{person}{Rumman
  Chowdhury}.} \bibinfo{year}{2021}\natexlab{}.
\newblock \showarticletitle{Where responsible AI meets reality: Practitioner
  perspectives on enablers for shifting organizational practices}.
\newblock \bibinfo{journal}{\emph{Proceedings of the ACM on Human-Computer
  Interaction}} \bibinfo{volume}{5}, \bibinfo{number}{CSCW1}
  (\bibinfo{year}{2021}), \bibinfo{pages}{1--23}.
\newblock


\bibitem[Research(2022)]%
        {MSRchecklist}
\bibfield{author}{\bibinfo{person}{Microsoft Research}.}
  \bibinfo{year}{2022}\natexlab{}.
\newblock \bibinfo{title}{AI Fairness Checklist}.
\newblock
\newblock
\urldef\tempurl%
\url{https://blog.google/technology/ai/join-us-in-the-ai-test-kitchen/}
\showURL{%
\tempurl}


\bibitem[Research(2021)]%
        {googlePAIR}
\bibfield{author}{\bibinfo{person}{People +~AI Research}.}
  \bibinfo{year}{2021}\natexlab{}.
\newblock \showarticletitle{People AI Guidebook}.
\newblock  (\bibinfo{year}{2021}).
\newblock
\urldef\tempurl%
\url{https://pair.withgoogle.com/guidebook/}
\showURL{%
\tempurl}


\bibitem[Richardson et~al\mbox{.}(2021)]%
        {richardson2021towards}
\bibfield{author}{\bibinfo{person}{Brianna Richardson}, \bibinfo{person}{Jean
  Garcia-Gathright}, \bibinfo{person}{Samuel~F Way}, \bibinfo{person}{Jennifer
  Thom}, {and} \bibinfo{person}{Henriette Cramer}.}
  \bibinfo{year}{2021}\natexlab{}.
\newblock \showarticletitle{Towards Fairness in Practice: A
  Practitioner-Oriented Rubric for Evaluating Fair ML Toolkits}. In
  \bibinfo{booktitle}{\emph{Proceedings of the 2021 CHI Conference on Human
  Factors in Computing Systems}}. \bibinfo{pages}{1--13}.
\newblock


\bibitem[Robertson and Salehi(2020)]%
        {robertson2020-what}
\bibfield{author}{\bibinfo{person}{Samantha Robertson} {and}
  \bibinfo{person}{Niloufar Salehi}.} \bibinfo{year}{2020}\natexlab{}.
\newblock \showarticletitle{What If I Don't Like Any Of The Choices? The Limits
  of Preference Elicitation for Participatory Algorithm Design}. In
  \bibinfo{booktitle}{\emph{Workshop on Participatory Approaches to Machine
  Learning at ICML 2020}}.
\newblock


\bibitem[Russakovsky et~al\mbox{.}(2015)]%
        {russakovsky2015imagenet}
\bibfield{author}{\bibinfo{person}{Olga Russakovsky}, \bibinfo{person}{Jia
  Deng}, \bibinfo{person}{Hao Su}, \bibinfo{person}{Jonathan Krause},
  \bibinfo{person}{Sanjeev Satheesh}, \bibinfo{person}{Sean Ma},
  \bibinfo{person}{Zhiheng Huang}, \bibinfo{person}{Andrej Karpathy},
  \bibinfo{person}{Aditya Khosla}, \bibinfo{person}{Michael Bernstein},
  {et~al\mbox{.}}} \bibinfo{year}{2015}\natexlab{}.
\newblock \showarticletitle{Imagenet large scale visual recognition challenge}.
\newblock \bibinfo{journal}{\emph{International journal of computer vision}}
  \bibinfo{volume}{115}, \bibinfo{number}{3} (\bibinfo{year}{2015}),
  \bibinfo{pages}{211--252}.
\newblock


\bibitem[Salehi et~al\mbox{.}(2015)]%
        {salehi2015we}
\bibfield{author}{\bibinfo{person}{Niloufar Salehi}, \bibinfo{person}{Lilly~C
  Irani}, \bibinfo{person}{Michael~S Bernstein}, \bibinfo{person}{Ali
  Alkhatib}, \bibinfo{person}{Eva Ogbe}, {and} \bibinfo{person}{Kristy
  Milland}.} \bibinfo{year}{2015}\natexlab{}.
\newblock \showarticletitle{We are dynamo: Overcoming stalling and friction in
  collective action for crowd workers}. In
  \bibinfo{booktitle}{\emph{Proceedings of the 33rd annual ACM conference on
  human factors in computing systems}}. \bibinfo{pages}{1621--1630}.
\newblock


\bibitem[Sandvig et~al\mbox{.}(2014)]%
        {sandvig2014auditing}
\bibfield{author}{\bibinfo{person}{Christian Sandvig}, \bibinfo{person}{Kevin
  Hamilton}, \bibinfo{person}{Karrie Karahalios}, {and} \bibinfo{person}{Cedric
  Langbort}.} \bibinfo{year}{2014}\natexlab{}.
\newblock \showarticletitle{Auditing algorithms: Research methods for detecting
  discrimination on internet platforms}.
\newblock \bibinfo{journal}{\emph{Data and Discrimination: Converting Critical
  Concerns into Productive Inquiry}} (\bibinfo{year}{2014}).
\newblock


\bibitem[Schwartz(2019)]%
        {Tay_Ban}
\bibfield{author}{\bibinfo{person}{Oscar Schwartz}.}
  \bibinfo{year}{2019}\natexlab{}.
\newblock \bibinfo{title}{Microsoft’s Racist Chatbot Revealed the Dangers of
  Online Conversation The bot learned language from people on Twitter—but it
  also learned values}.
\newblock
\newblock
\urldef\tempurl%
\url{https://spectrum.ieee.org/in-2016-microsofts-racist-chatbot-revealed-the-dangers-of-online-conversation}
\showURL{%
\tempurl}


\bibitem[Selbst et~al\mbox{.}(2019)]%
        {selbst2019fairness}
\bibfield{author}{\bibinfo{person}{Andrew~D. Selbst}, \bibinfo{person}{Danah
  Boyd}, \bibinfo{person}{Sorelle~A. Friedler}, \bibinfo{person}{Suresh
  Venkatasubramanian}, {and} \bibinfo{person}{Janet Vertesi}.}
  \bibinfo{year}{2019}\natexlab{}.
\newblock \showarticletitle{Fairness and Abstraction in Sociotechnical
  Systems}.
\newblock  (\bibinfo{year}{2019}), \bibinfo{pages}{59–68}.
\newblock
\showISBNx{9781450361255}
\urldef\tempurl%
\url{https://doi.org/10.1145/3287560.3287598}
\showDOI{\tempurl}


\bibitem[Shen et~al\mbox{.}(2021)]%
        {shen2021everyday}
\bibfield{author}{\bibinfo{person}{Hong Shen}, \bibinfo{person}{Alicia DeVos},
  \bibinfo{person}{Motahhare Eslami}, {and} \bibinfo{person}{Kenneth
  Holstein}.} \bibinfo{year}{2021}\natexlab{}.
\newblock \showarticletitle{Everyday algorithm auditing: Understanding the
  power of everyday users in surfacing harmful algorithmic behaviors}.
\newblock \bibinfo{journal}{\emph{Proceedings of the ACM on Human-Computer
  Interaction}} \bibinfo{volume}{5}, \bibinfo{number}{CSCW2}
  (\bibinfo{year}{2021}), \bibinfo{pages}{1--29}.
\newblock


\bibitem[Shen et~al\mbox{.}(2022)]%
        {shen2022model}
\bibfield{author}{\bibinfo{person}{Hong Shen}, \bibinfo{person}{Leijie Wang},
  \bibinfo{person}{Wesley~H Deng}, \bibinfo{person}{Ciell Brusse},
  \bibinfo{person}{Ronald Velgersdijk}, {and} \bibinfo{person}{Haiyi Zhu}.}
  \bibinfo{year}{2022}\natexlab{}.
\newblock \showarticletitle{The Model Card Authoring Toolkit: Toward
  Community-centered, Deliberation-driven AI Design}. In
  \bibinfo{booktitle}{\emph{2022 ACM Conference on Fairness, Accountability,
  and Transparency}}. \bibinfo{pages}{440--451}.
\newblock


\bibitem[Sloane et~al\mbox{.}(2020)]%
        {sloane2020participation}
\bibfield{author}{\bibinfo{person}{Mona Sloane}, \bibinfo{person}{Emanuel
  Moss}, \bibinfo{person}{Olaitan Awomolo}, {and} \bibinfo{person}{Laura
  Forlano}.} \bibinfo{year}{2020}\natexlab{}.
\newblock \showarticletitle{Participation is not a design fix for machine
  learning}.
\newblock \bibinfo{journal}{\emph{arXiv preprint arXiv:2007.02423}}
  (\bibinfo{year}{2020}).
\newblock


\bibitem[Snow et~al\mbox{.}(2008)]%
        {snow2008cheap}
\bibfield{author}{\bibinfo{person}{Rion Snow}, \bibinfo{person}{Brendan
  O’connor}, \bibinfo{person}{Dan Jurafsky}, {and} \bibinfo{person}{Andrew~Y
  Ng}.} \bibinfo{year}{2008}\natexlab{}.
\newblock \showarticletitle{Cheap and fast--but is it good? evaluating
  non-expert annotations for natural language tasks}. In
  \bibinfo{booktitle}{\emph{Proceedings of the 2008 conference on empirical
  methods in natural language processing}}. \bibinfo{pages}{254--263}.
\newblock


\bibitem[Steiger et~al\mbox{.}(2021)]%
        {steiger2021psychological}
\bibfield{author}{\bibinfo{person}{Miriah Steiger}, \bibinfo{person}{Timir~J
  Bharucha}, \bibinfo{person}{Sukrit Venkatagiri}, \bibinfo{person}{Martin~J
  Riedl}, {and} \bibinfo{person}{Matthew Lease}.}
  \bibinfo{year}{2021}\natexlab{}.
\newblock \showarticletitle{The psychological well-being of content moderators:
  the emotional labor of commercial moderation and avenues for improving
  support}. In \bibinfo{booktitle}{\emph{Proceedings of the 2021 CHI conference
  on human factors in computing systems}}. \bibinfo{pages}{1--14}.
\newblock


\bibitem[Suh et~al\mbox{.}(2019)]%
        {suh2019anchorviz}
\bibfield{author}{\bibinfo{person}{Jina Suh}, \bibinfo{person}{Soroush
  Ghorashi}, \bibinfo{person}{Gonzalo Ramos}, \bibinfo{person}{Nan-Chen Chen},
  \bibinfo{person}{Steven Drucker}, \bibinfo{person}{Johan Verwey}, {and}
  \bibinfo{person}{Patrice Simard}.} \bibinfo{year}{2019}\natexlab{}.
\newblock \showarticletitle{AnchorViz: Facilitating Semantic Data Exploration
  and Concept Discovery for Interactive Machine Learning}.
\newblock \bibinfo{journal}{\emph{ACM Trans. Interact. Intell. Syst.}}
  \bibinfo{volume}{10}, \bibinfo{number}{1}, Article \bibinfo{articleno}{7}
  (\bibinfo{date}{Aug.} \bibinfo{year}{2019}), \bibinfo{numpages}{38}~pages.
\newblock
\showISSN{2160-6455}
\urldef\tempurl%
\url{https://doi.org/10.1145/3241379}
\showDOI{\tempurl}


\bibitem[Sweeney(2013)]%
        {sweeney2013discrimination}
\bibfield{author}{\bibinfo{person}{Latanya Sweeney}.}
  \bibinfo{year}{2013}\natexlab{}.
\newblock \showarticletitle{Discrimination in online ad delivery}.
\newblock \bibinfo{journal}{\emph{Queue}} \bibinfo{volume}{11},
  \bibinfo{number}{3} (\bibinfo{year}{2013}), \bibinfo{pages}{10--29}.
\newblock


\bibitem[Varshney(2019)]%
        {varshney2019trustworthy}
\bibfield{author}{\bibinfo{person}{Kush~R Varshney}.}
  \bibinfo{year}{2019}\natexlab{}.
\newblock \showarticletitle{Trustworthy machine learning and artificial
  intelligence}.
\newblock \bibinfo{journal}{\emph{XRDS: Crossroads, The ACM Magazine for
  Students}} \bibinfo{volume}{25}, \bibinfo{number}{3} (\bibinfo{year}{2019}),
  \bibinfo{pages}{26--29}.
\newblock


\bibitem[Vaughan(2017)]%
        {vaughan2017making}
\bibfield{author}{\bibinfo{person}{Jennifer~Wortman Vaughan}.}
  \bibinfo{year}{2017}\natexlab{}.
\newblock \showarticletitle{Making Better Use of the Crowd: How Crowdsourcing
  Can Advance Machine Learning Research.}
\newblock \bibinfo{journal}{\emph{J. Mach. Learn. Res.}} \bibinfo{volume}{18},
  \bibinfo{number}{1} (\bibinfo{year}{2017}), \bibinfo{pages}{7026--7071}.
\newblock


\bibitem[Veale and Binns(2017)]%
        {veale2017fairer}
\bibfield{author}{\bibinfo{person}{Michael Veale} {and} \bibinfo{person}{Reuben
  Binns}.} \bibinfo{year}{2017}\natexlab{}.
\newblock \showarticletitle{Fairer machine learning in the real world:
  Mitigating discrimination without collecting sensitive data}.
\newblock \bibinfo{journal}{\emph{Big Data \& Society}} \bibinfo{volume}{4},
  \bibinfo{number}{2} (\bibinfo{year}{2017}),
  \bibinfo{pages}{2053951717743530}.
\newblock


\bibitem[Warkentin and Woodward(2022)]%
        {AItest}
\bibfield{author}{\bibinfo{person}{Tris Warkentin} {and} \bibinfo{person}{Josh
  Woodward}.} \bibinfo{year}{2022}\natexlab{}.
\newblock \bibinfo{title}{AI Test Kitchen}.
\newblock
\newblock
\urldef\tempurl%
\url{https://blog.google/technology/ai/join-us-in-the-ai-test-kitchen/}
\showURL{%
\tempurl}


\bibitem[Wu et~al\mbox{.}(2019)]%
        {wu2019errudite}
\bibfield{author}{\bibinfo{person}{Tongshuang Wu}, \bibinfo{person}{Marco~Tulio
  Ribeiro}, \bibinfo{person}{Jeffrey Heer}, {and} \bibinfo{person}{Daniel~S
  Weld}.} \bibinfo{year}{2019}\natexlab{}.
\newblock \showarticletitle{Errudite: Scalable, reproducible, and testable
  error analysis}. In \bibinfo{booktitle}{\emph{Proceedings of the 57th Annual
  Meeting of the Association for Computational Linguistics}}.
  \bibinfo{pages}{747--763}.
\newblock


\bibitem[Young et~al\mbox{.}(2019)]%
        {young2019toward}
\bibfield{author}{\bibinfo{person}{Meg Young}, \bibinfo{person}{Lassana
  Magassa}, {and} \bibinfo{person}{Batya Friedman}.}
  \bibinfo{year}{2019}\natexlab{}.
\newblock \showarticletitle{Toward inclusive tech policy design: a method for
  underrepresented voices to strengthen tech policy documents}.
\newblock \bibinfo{journal}{\emph{Ethics and Information Technology}}
  \bibinfo{volume}{21}, \bibinfo{number}{2} (\bibinfo{year}{2019}),
  \bibinfo{pages}{89--103}.
\newblock


\bibitem[Zaidan and Callison-Burch(2011)]%
        {zaidan2011crowdsourcing}
\bibfield{author}{\bibinfo{person}{Omar Zaidan} {and} \bibinfo{person}{Chris
  Callison-Burch}.} \bibinfo{year}{2011}\natexlab{}.
\newblock \showarticletitle{Crowdsourcing translation: Professional quality
  from non-professionals}. In \bibinfo{booktitle}{\emph{Proceedings of the 49th
  annual meeting of the association for computational linguistics: human
  language technologies}}. \bibinfo{pages}{1220--1229}.
\newblock


\bibitem[Zhou et~al\mbox{.}(2019)]%
        {zhou2019hype}
\bibfield{author}{\bibinfo{person}{Sharon Zhou}, \bibinfo{person}{Mitchell
  Gordon}, \bibinfo{person}{Ranjay Krishna}, \bibinfo{person}{Austin Narcomey},
  \bibinfo{person}{Li~F Fei-Fei}, {and} \bibinfo{person}{Michael Bernstein}.}
  \bibinfo{year}{2019}\natexlab{}.
\newblock \showarticletitle{Hype: A benchmark for human eye perceptual
  evaluation of generative models}.
\newblock \bibinfo{journal}{\emph{Advances in neural information processing
  systems}}  \bibinfo{volume}{32} (\bibinfo{year}{2019}).
\newblock


\bibitem[Zou and Schiebinger(2018)]%
        {zou2018ai}
\bibfield{author}{\bibinfo{person}{James Zou} {and} \bibinfo{person}{Londa
  Schiebinger}.} \bibinfo{year}{2018}\natexlab{}.
\newblock \showarticletitle{AI can be sexist and racist—it’s time to make
  it fair}.
\newblock \bibinfo{journal}{\emph{Nature}}  \bibinfo{volume}{559}
  (\bibinfo{year}{2018}), \bibinfo{pages}{324--326}.
\newblock


\end{thebibliography}
\clearpage
\onecolumn
\appendix

\section{Appendix}

\begin{figure*}[h]
  \centering
  \includegraphics[width=0.690\linewidth]{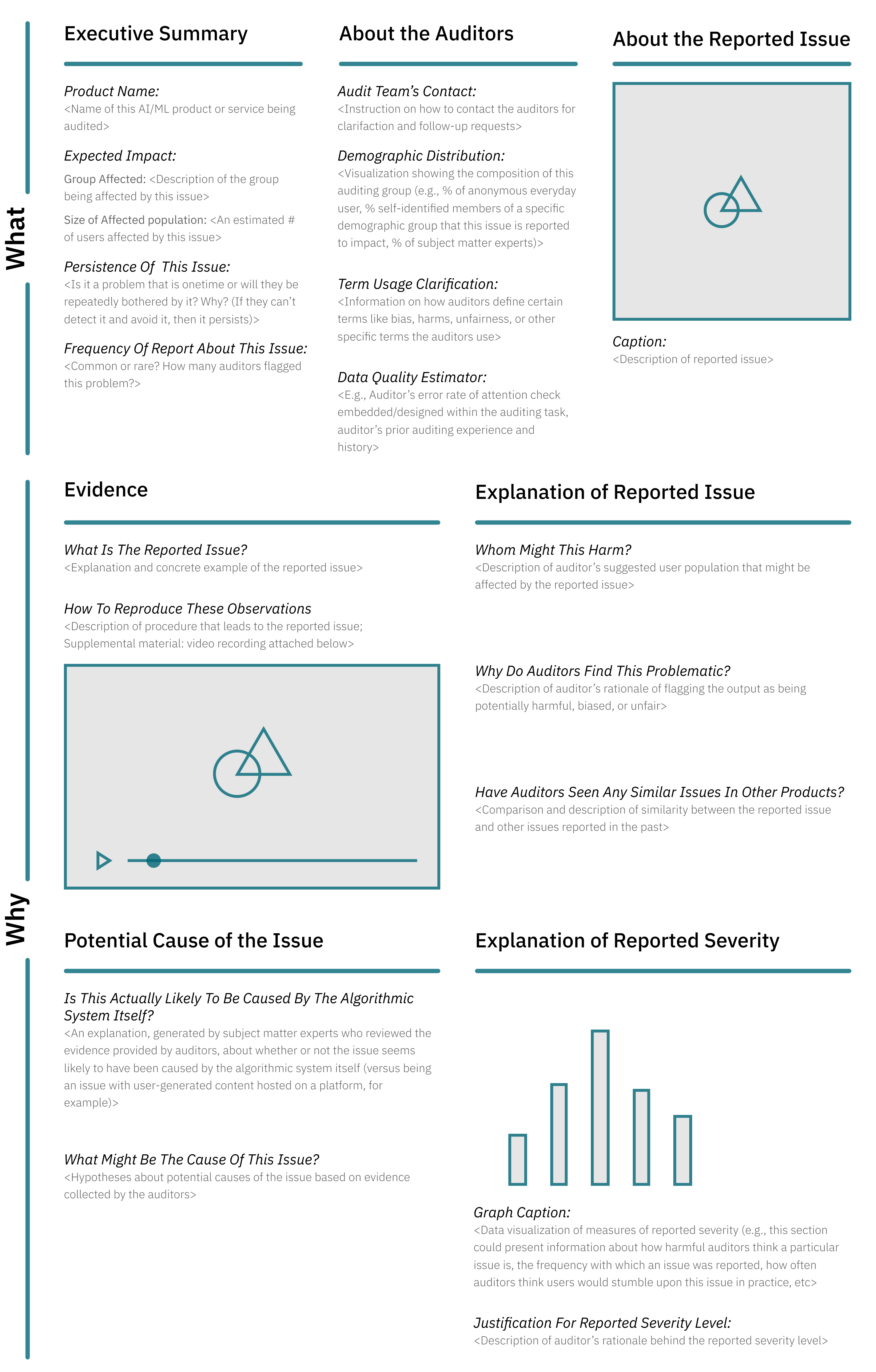}
  \caption[]{
    A potential \rr{user-engaged} audit report that was iteratively co-designed with participants. During the co-design activity, participants could zoom in, annotate, and modify the details. We used this report template as a \textbf{probe}, \textbf{not as final products}, to investigate more deeply on practitioners’ challenges and desires.
}
  \Description{Potential user-engaged audit report that were iteratively co-designed with participants. Each section includes different information desired by AI practitioners, such as auditor's information, explanation of reported issues, and potential cause of the issue.}
  \label{fig:report}
\end{figure*}


\end{document}